\documentclass[pdflatex,sn-mathphys-num]{sn-jnl}% Math and Physical Sciences Numbered Reference Style 

\usepackage{graphicx}
\usepackage{multirow}
\usepackage{amsmath,amssymb,amsfonts}
\usepackage[title]{appendix}
\usepackage{xcolor}
\usepackage{lineno}
\usepackage{xspace}
\usepackage{subcaption}

\newcommand{\hess}{H.E.S.S.\xspace}
\newcommand{\fmlat}{\emph{Fermi}-LAT\xspace}
\newcommand{\hone}{H~{\sc i}\xspace}
\newcommand{\hii}{H~{\sc ii}\xspace}
\newcommand{\wld}{Westerlund~1\xspace}
\newcommand{\outflow}{J1654$-$467\xspace}

\newcommand{\aap}{Astronomy \& Astrophysics}
\newcommand{\aj}{Astronomical Journal}
\newcommand{\apj}{Astrophysical Journal}
\newcommand{\apjl}{Astrophysical Journal Letters}
\newcommand{\apjs}{Astrophysical Journal Supplement Series}
\newcommand{\mnras}{Monthly Notices of the Royal Astronomical Society}
\newcommand{\araa}{Annual Review of Astronomy and Astrophysics}
\newcommand{\aapr}{Astronomy and Astrophysics Reviews}
\newcommand{\pasa}{Publications of the Astronomical Society of Australia}

\unnumbered

\begin{document}

\title{A cosmic-ray loaded nascent outflow driven by a massive star cluster}

\author[1]{\fnm{Marianne} \sur{Lemoine-Goumard}}\email{lemoine@cenbg.in2p3.fr}
\equalcont{These authors contributed equally to this work.}

\author[2]{\fnm{Lucia} \sur{H\"arer}}\email{lucia.haerer@mpi-hd.mpg.de}
\equalcont{These authors contributed equally to this work.}

\author[2]{\fnm{Lars} \sur{Mohrmann}}\email{lars.mohrmann@mpi-hd.mpg.de}
\equalcont{These authors contributed equally to this work.}

\author[3,4]{\fnm{Romain} \sur{Bernet}} %\email{Romain.BERNET@student.isae-supaero.fr}

\author[2]{\fnm{Jim} \sur{Hinton}} %\email{jim.hinton@mpi-hd.mpg.de}

\author[5]{\fnm{Giada} \sur{Peron}}%\email{giada.peron@inaf.it }

\author[2]{\fnm{Brian} \sur{Reville}}%\email{brian.reville@mpi-hd.mpg.de}

\author[3]{\fnm{Luigi} \sur{Tibaldo}}%\email{luigi.tibaldo@irap.omp.eu}

\author[2]{\fnm{Thibault} \sur{Vieu}}%\email{thibault.vieu@mpi-hd.mpg.de}

\affil[1]{Universit\'e Bordeaux, CNRS, LP2I Bordeaux, UMR 5797, F-33170 Gradignan, France}

\affil[2]{Max-Planck-Institut f\"ur Kernphysik, Saupfercheckweg 1, D-69117 Heidelberg, Germany}

\affil[3]{IRAP, Universit\'e de Toulouse, CNRS, CNES, UPS, F-31028 Toulouse, France}

\affil[4]{Institut Sup\'erieur de l'A\'eronautique et de l'Espace (ISAE-SUPAERO), Universit\'e de Toulouse, F-31055 Toulouse, France}

\affil[5]{INAF Osservatorio Astrofisico Arcetri, Largo E. Fermi, 5, 50125, Florence, Italy}

%\linenumbers

% \abstract{}

% \keywords{Keyword1, Keyword2, Keyword3, Keyword4}

\maketitle

{\bf
Cosmic rays are widely held to drive outflows from star-forming galaxies and profoundly influence galaxy evolution.
Direct evidence for cosmic-ray carrying outflows is however lacking.
At the same time there is increasing awareness of the importance of massive star clusters in the acceleration of cosmic rays in galaxies.
Here we report on the discovery of a nascent outflow driven by the massive star cluster \wld.
Giga-electronvolt gamma-ray emission coincident with a cavity visible in atomic hydrogen traces the emergence of a population of relativistic electrons out of the Galactic Disc.
The emission is offset from tera-electronvolt gamma-ray radiation surrounding the cluster, but connects to it smoothly spectrally and spatially.
The implied energy density of co-accelerated protons and nuclei, assuming standard non-thermal electron/proton injection efficiencies, is at least an order of magnitude higher than that in the general interstellar medium.
These particles therefore have the potential to dynamically influence the outflow.
This discovery suggests that cosmic-ray loaded outflows may be a common feature of young massive star clusters, with implications for the transport of cosmic rays into the halo of the Galaxy.
}

\bigskip

\subsection*{Introduction}

Observations of nearby starburst galaxies suggest the presence of global cosmic-ray (CR) pressure driven winds \cite{ThompsonHeckman}.
It is unclear however whether the CR content in the halo of the Milky Way is sufficient to accelerate such a wind.
On the other hand, galaxy-scale simulations are in agreement that CR feedback at high latitudes succeeds in driving large-scale winds \cite{Girichidis, Rathjen2021, Modak,Armilotta,Sike2024,Kjellgren2025}.
Since these winds affect the nearby circumgalactic medium and, as a consequence, the Galactic star-formation rate, the CR content in the Galactic halo and the processes that regulate how CRs get there are critical components of any model of the interstellar medium (ISM) \cite{Ruszkowski}. 

Early fluid \cite{JohnsonAxford,Ipavich} and kinetic \cite{BreitschwerdtI,BreitschwerdtII} models predicted the existence of Galactic winds driven by CRs in the halo.
Modern simulations -- encompassing stellar feedback, multiple supernovae, the resulting CR production, and interaction with a multi-phase ISM -- can now present a more consistent picture of this process \cite[e.g.][]{Girichidis,Rathjen2021, Modak,Armilotta,Sike2024,Kjellgren2025}.
These works find that the kpc-scale outflows are sensitive not only to the gas conditions in and around superbubbles, but also to the assumptions on CR transport and associated plasma effects.
These effects, such as ion-neutral damping, depend upon the phase of the surrounding ISM, and in particular that above and below the Galactic Plane.

The material feeding these halo winds originates from stellar activity, especially supernovae \cite{McKee1977,Ferriere}, which are unevenly distributed throughout the disk.
To reach the halo, material must penetrate the diffuse warm ISM (WIM; see \cite{MacLow1989} for further details).
The WIM density falls off exponentially away from the mid-plane, with a scale-height of $\approx 1$kpc \cite{Gaensler08}, substantially longer than that of the thin disc of the Galactic Plane ($\approx 100$\,pc \cite{Ferriere}).
Here young massive star clusters (YMSCs) have a distinct advantage. The combined action of powerful winds and clustering of supernovae in the first 10 Myr of the YMSC's evolution are known to inflate superbubbles, that can extend over several hundreds of pc, exceeding the scale-height of the Galactic Plane.
Due to the gradient in external gas density normal to the Galactic Disc, the bubbles can become asymmetric, and expand preferentially down the gradient forming structures called chimneys \cite{Norman1989, Ponti2019}.

The connection between superbubble-driven outflows and CRs is especially significant, given the growing interest in YMSCs as CR sources \cite{Aharonian2019}, and in particular their contribution to Galactic CRs at the highest energies \cite{Morlino2021,Vieu2023}.
Gamma-ray observations of YMSCs at giga-electronvolt (GeV) energies \cite[e.g.][]{FermiLAT_Cocoon_2011,Yang2017,Yang2020,Sun2020,Liu2022,Peron2024} as well as tera-electronvolt (TeV) energies \cite{HESS_Wd2_2011,HAWC_Cygnus_2021,HESS_Wd1_2022,HESS_LMC_2024,LHAASO_Cygnus_2024,LHAASO_W43_2025} indicate they are sites of efficient CR production.
Gamma-ray observations can therefore be used as probes of superbubble-driven outflows, constraining the entrained CR content therein. 

\wld, as the most massive YMSC observed in our Galaxy, offers an ideal target for gamma-ray studies.
Its cluster core is located at a distance of about 4\,kpc \cite{Navarete2022,Negueruela2022} and is known to host a wealth of massive stars \cite{Clark2005,Crowther2006}.
Multiple estimates place the age of the cluster at around 4\,Myr \cite{Clark2005,Crowther2006,Brandner2008}, although recent studies have found evidence for sub-populations of stars with ages of up to 10\,Myr \cite{Beasor2021,Navarete2022}.
Within the last Myr, the kinetic wind-power output of the cluster has plausibly been of the order of $10^{39}\,\mathrm{erg}\,\mathrm{s}^{-1}$ \cite{Muno2006}.

\wld is unique among YMSCs in that it is surrounded by a resolved gamma-ray ``ring'' at TeV energies, spanning radii of about 20--50\,pc \cite{HESS_Wd1_2022}.
This ring is thought to be connected with the superbubble around the cluster and has been hypothesised to arise from inverse-Compton (IC) emission of high-energy electrons accelerated at the termination shock of the collective cluster wind \cite{Haerer2023}.
CR protons and nuclei (``hadronic CRs''), although almost inevitably produced alongside electrons, produce effectively measurable gamma-ray emission only in the presence of relatively dense target material.
The hot, low-density gas expected in an outflow from \wld is therefore not favourable for hadronic gamma-ray production.
The accelerated electrons, on the other hand, will inevitably cool via inverse-Compton scattering, with a large fraction of the available power converted to gamma rays.
As the most massive YMSC observed in the Galaxy, with established ongoing CR acceleration to beyond $100$ TeV, \wld presents the most promising target to understand the propagation of accelerated CRs away from their sources, and probe the emergence of a large-scale outflow. 
To probe the system as a whole requires electrons with cooling times comparable to the age of the cluster itself, corresponding to energies of approximately 30\,GeV (see Methods) and resulting gamma-ray emission at around 3 GeV, in the range of the \fmlat instrument \cite{FermiLAT2009}.

Here we show that the electron population that produces the TeV gamma-ray emission connects smoothly to the expected large scale outflow emerging below the Galactic Plane.
The entrained electrons are revealed by their inverse-Compton emission which we detect with the \fmlat instrument.
The presence of such an outflow is consistent with a low-density feature seen in relevant gas maps. These findings indicate the presence of a nascent cosmic-ray loaded Galactic outflow.

\subsection*{Results and Discussion}
\subsubsection*{\fmlat data analysis}
\label{sec:data}
Using 15 years of \fmlat data, we carried out a deep morphological and spectral analysis of the gamma-ray emission in a $15^{\circ} \times 15^{\circ}$ region around \wld in the 3\,GeV -- 3\,TeV energy band.
As our aim is to connect with the \hess measurements of the \wld region, and to reduce uncertainties related to the modelling of diffuse gamma-ray emission, we do not consider gamma rays below 3\,GeV in this work.
We put special emphasis on accurately modelling the diffuse interstellar radiation arising from interactions of the Galactic CR sea with gas and magnetic fields (see Methods).
% Full details are provided in Appendix~\ref{sec:appx_fermi_lat}.

\begin{figure}
  \centering
  \includegraphics[width=0.7\textwidth]{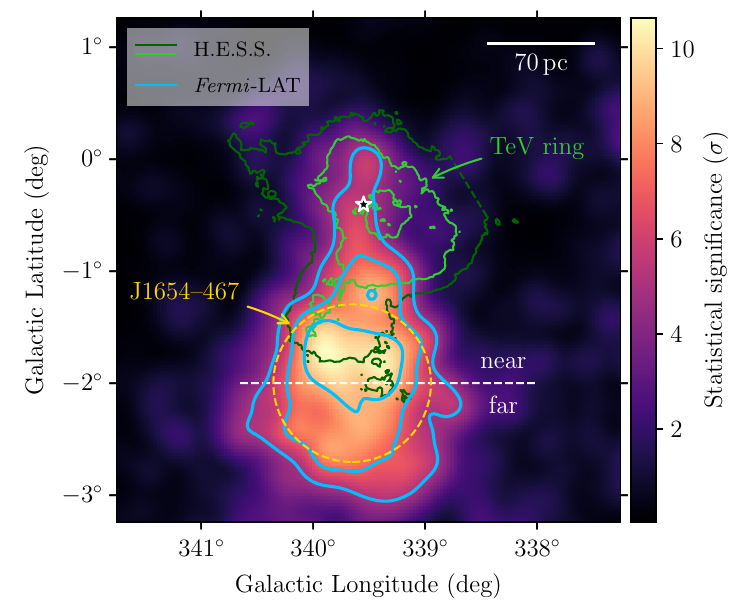}
  \caption{
    \textbf{\fmlat test statistic (TS) map.}
    The colour scale displays $\sqrt{\mathrm{TS}}$, i.e. the statistical significance of a point-like source with a spectrum $\propto E^{-2}$, in the energy range 3\,GeV--3\,TeV.
    Contour lines at $\sqrt{\mathrm{TS}}=(5, 7, 9)$ are shown in blue.
    The orange, dashed circle denotes the 1-$\sigma$ radius of the Gaussian source model for \outflow.
    Flux contours of the TeV gamma-ray emission measured with \hess \cite{HESS_Wd1_2022} at levels of $(1.1/2.3)\times 10^{-8}\,\mathrm{cm}^{-2}\,\mathrm{s}^{-1}\,\mathrm{sr}^{-1}$ are shown in dark and light green, respectively.
    The star marker indicates the position of \wld.
    The white, dashed line separates the \outflow region into two halves, for which we derive separate spectra (cf.\ Fig.~\ref{fig:sed}).
    The scale bar assumes a distance to \wld of 4.14\,kpc \cite{Negueruela2022,Navarete2022}.
  }
  \label{fig:fermi_ts_map}
\end{figure}

The result of the \fmlat analysis is illustrated in Fig.~\ref{fig:fermi_ts_map}.
Our study reveals two new components with respect to previous studies:
(i)~an emission region coincident with \wld that we model with a template derived from the \hess TeV flux map \cite{HESS_Wd1_2022};
(ii)~a diffuse component extending from the cluster location away from the Galactic Plane, best represented by a Gaussian source ($l=(339.61 \pm  0.03)^\circ$, $b=(-1.91 \pm 0.04)^\circ$, $\sigma = (0.71 \pm 0.03)^\circ$; shown by the dashed orange circle in Fig.~\ref{fig:fermi_ts_map}).
We will refer to this second new component as \outflow.

% As we demonstrate in appendices~\ref{sec:appx_fermi_lat} and~\ref{appx:hess_fermi_sed_comparison}, the first component represents a counterpart to the ring-like TeV gamma-ray emission detected with \hess, being compatible both in terms of spatial morphology and energy spectra.
As we demonstrate in the Methods section, the first component represents a counterpart to the ring-like TeV gamma-ray emission detected with \hess, being compatible both in terms of spatial morphology and energy spectra.
\outflow, on the other hand, appears to be a continuation of the TeV gamma-ray radiation detected with \hess, which features emission protruding from the ring in the same direction (named HESS~J1652$-$462 in ref.~\cite{HESS_Wd1_2022}).
The structure is approximately 150\,pc long in projection and, with a gamma-ray energy flux above 10 GeV of $(10.5 \pm 0.5_\mathrm{stat} \pm 0.4_\mathrm{syst}) \times 10^{-11}\,\mathrm{erg}\,\mathrm{cm}^{-2}\,\mathrm{s}^{-1}$, is the dominant source of 10--100\,GeV photons in the region.
That is in contrast to the picture at TeV energies, where the ``annex'' is less extended and not as bright as the emission surrounding the star cluster.
This is naturally explained by the energy-dependent cooling rate of radiating electrons (see Methods).
In contrast, the indication of spectral softening with distance from \wld that we observe from the \fmlat data themselves (see Methods) is not explained by the difference in cooling time scales, but could be a feature of energy-dependent particle transport.
A search for possible multi-wavelength counterparts to \outflow did not reveal any candidates besides \wld.
In particular, we find that the two pulsars PSR~J1648$-$4611 and PSR~J1650$-$4601, whose positions on the sky are coincident with the southern part of the ``TeV ring'' -- while making a dominant contribution to the emission below 10\,GeV -- cannot be invoked to explain the presence of \outflow (see Methods).

\subsubsection*{Interstellar medium density}
\label{sec:ism-density}
\begin{figure}
  \centering
  \includegraphics[width=0.5\linewidth]{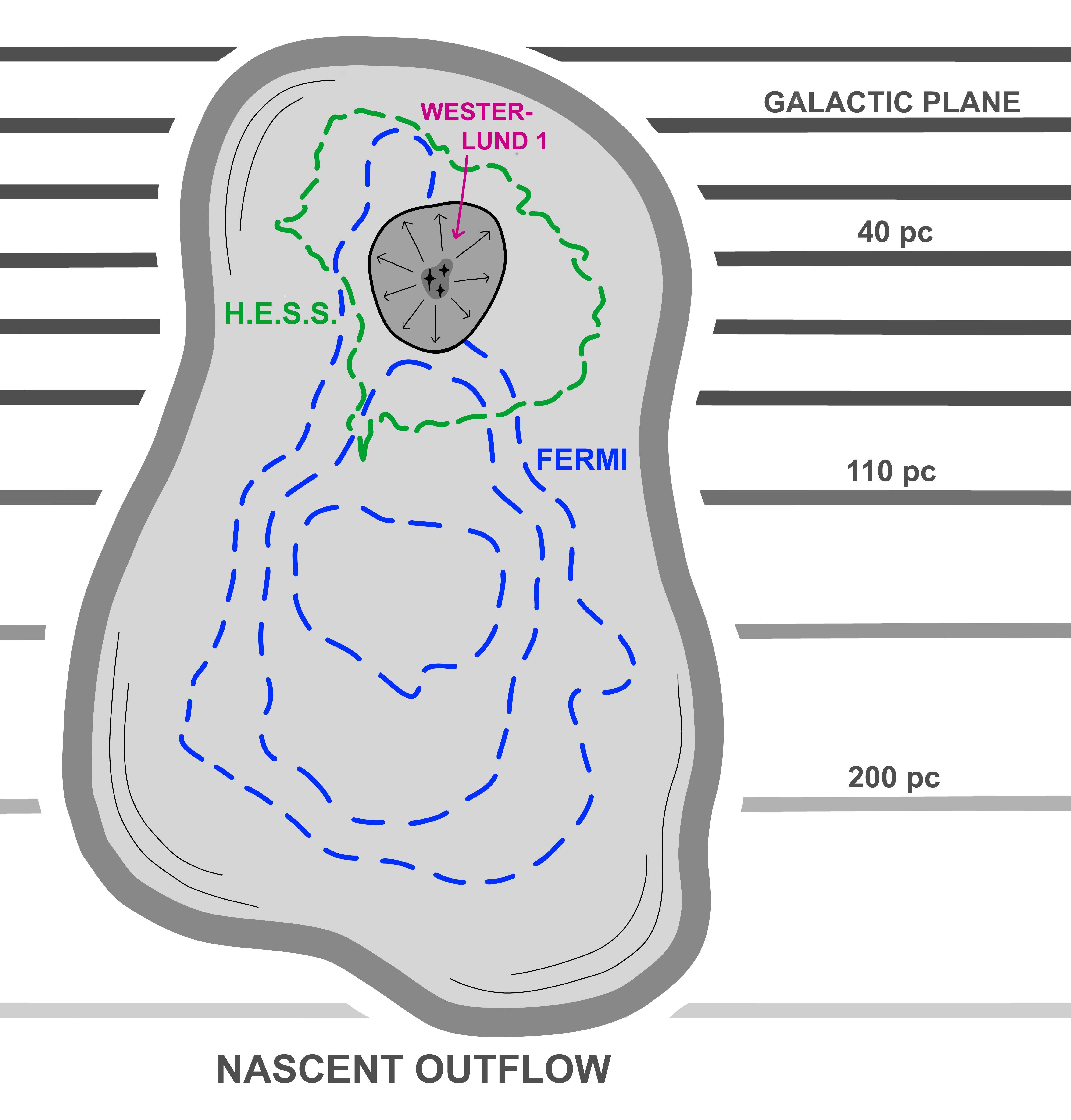}
  \caption{
    \textbf{Sketch of nascent outflow around \wld.}
    Massive stars and supernovae in \wld excavate material around the cluster, resulting in a low-density bubble, which expands asymmetrically due to the density gradient in the Galactic Disc.
    This will eventually lead to the formation of an open outflow connecting to the halo.
    Cosmic rays are accelerated at the cluster wind termination shock (black line).
    High-energy electrons exhibit short cooling times and generate the TeV-energy gamma rays measured with \hess (green contour \cite{HESS_Wd1_2022}); lower-energy electrons can travel further and are transported along the nascent outflow, where they are responsible for producing the GeV gamma-ray emission detected with \fmlat (blue contours).
    % The outer contour of the TeV-ring seen by \hess is shown in green and \fmlat contours in blue.
    Note that distances are not corrected for projection effects, i.e., the same scale as in Fig.~\ref{fig:fermi_ts_map} is used.
  }
  \label{fig:sketch}
\end{figure}

A natural explanation for the new structure is that it traces a population of relativistic particles accelerated at or near the cluster \wld that is breaking through the Galactic Plane in a nascent outflow (see Fig.~\ref{fig:sketch}).
Such outflows are expected for superbubble expansion in a stratified medium \cite{MacLow1989, Baumgartner}.

\begin{figure}
  \centering
  \includegraphics[width=\textwidth]{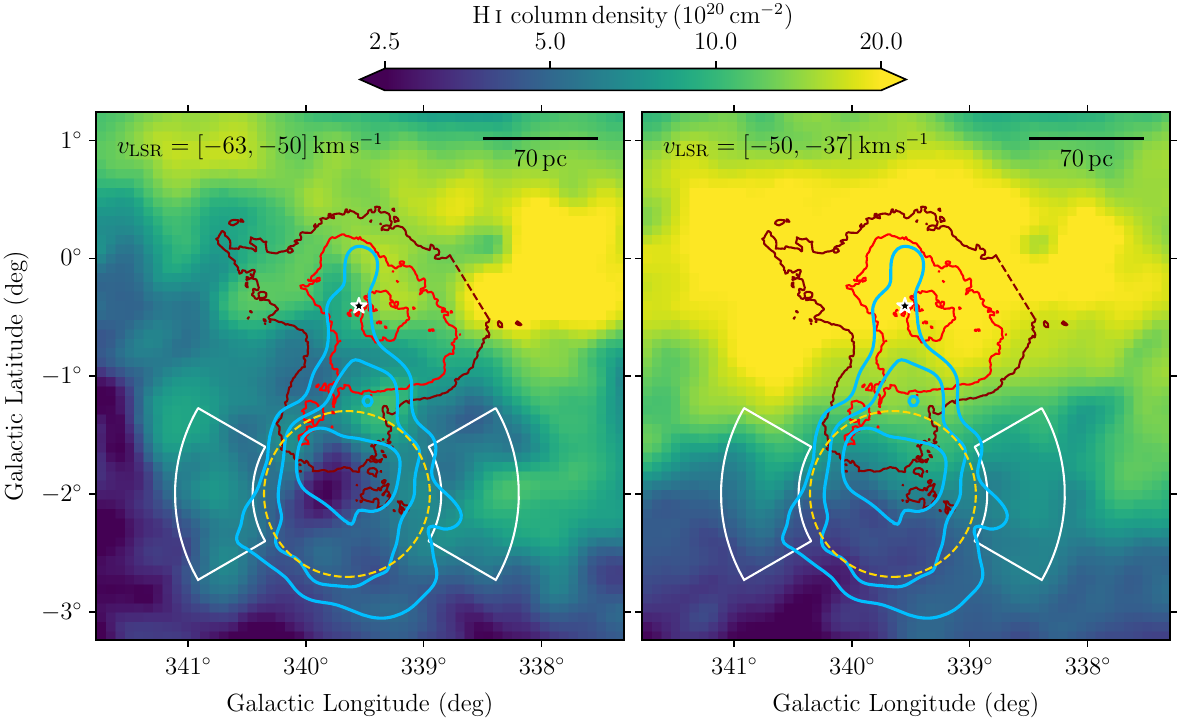}
  \caption{
    \textbf{Atomic (\hone) gas maps of the region around \wld.}
    Data are taken from the Parkes Galactic All-Sky Survey third data release (GASS~III) \cite{Kalberla2015}.
    The maps are derived using the optically thin approximation and are shown for the velocity ranges indicated in each panel.
    \fmlat GeV emission contours (blue), extent of \outflow (dashed orange), \hess TeV flux contours (here red), and position of \wld (star marker) are the same as in Fig.~\ref{fig:fermi_ts_map}.
    The white arc-shaped regions have been used to compute the difference in gas density between the outflow region and its surroundings.
  }
  \label{fig:gas_maps}
\end{figure}

This scenario can be tested by looking for an under-density, or cavity, in the Galactic gas emission that such a nascent outflow would produce.
Because no dense molecular hydrogen gas is present near \outflow (see Supplementary Figure 3), we focus here on atomic hydrogen gas.
Fig.~\ref{fig:gas_maps} shows \hone emission at velocities (with respect to the local standard of rest) around the Galactic rotation velocity of \wld ($v_\mathrm{LSR}\approx -50\,\mathrm{km}\,\mathrm{s}^{-1}$, see Methods).
A clear under-density is apparent over an extended velocity range, coincident with the peak \fmlat emission.
This under-density manifests itself in a difference in column density of approximately $(0.7-1.5)\times 10^{20}$\,cm$^{-2}$ with respect to neighbouring lines of sight.
Assuming that the depth of the structure is similar to its width, this corresponds to a deficit of about 0.3--0.7~hydrogen atoms\,cm$^{-3}$ (see Methods).
The energy required to inflate this structure depends on the assumed external temperature ($T$), the depth in projection ($d$) and the nature of the contents, but is of order $10^{50}\,(T/10^{4}\,\rm{K})(n/ 1\,\rm{cm}^{-3})(d/100\,\rm{pc})^3$\,erg, which is readily available given the  power of \wld.

\subsubsection*{Modelling}

\begin{figure}
  \centering
  \includegraphics[width=0.7\linewidth]{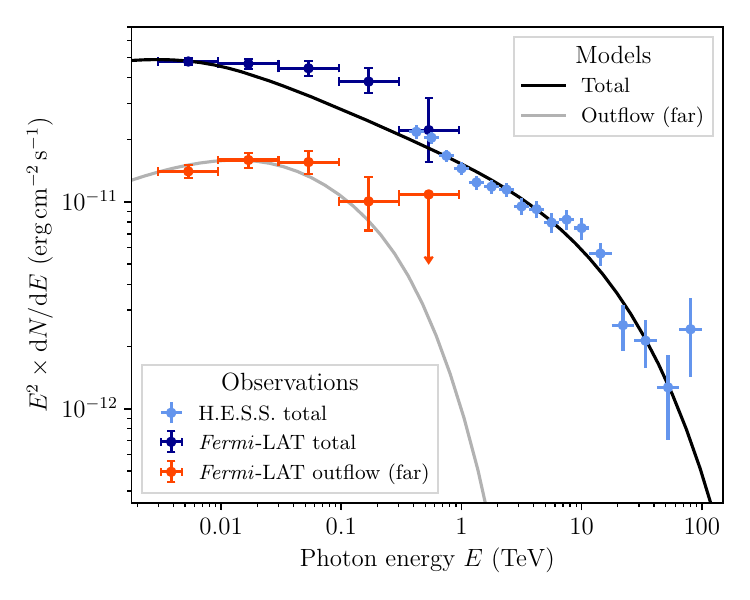}
  \caption{
	\textbf{Spectral energy distribution of the gamma-ray emission.}
	The dark and light blue points show the total flux measured with \fmlat (from the ``TeV ring'' and \outflow) and \hess, respectively, illustrating the spectral continuity despite the visually different spatial distributions.
	The black line shows a single-zone IC model for the whole population of electrons accelerated in the superbubble.
	The red points show the spectrum measured in the ``far'' region (cf.\ Fig.~\ref{fig:fermi_ts_map}).
    Error bars denote 68\% c.l.\ statistical uncertainties; upper limits are at 95\% c.l.
	The grey line follows from a simple time-dependent model for electrons in this region after 125\,kyr.
  }
  \label{fig:sed}
\end{figure}

%% summary of observational facts and TeV interpretation
The spectral energy distribution (SED), as presented in Fig.~\ref{fig:sed}, demonstrates that the total \fmlat spectrum connects smoothly to the \hess observed TeV gamma-ray emission surrounding \wld.
The same is found to hold true in different sub-regions (see Methods). This suggest a common origin of the emission from the entire region.  
As discussed above, a leptonic interpretation presents a natural solution to both the lack of correlation with gas that one would expect in a hadronic dominated scenario, and the energy-dependent morphology.
This interpretation complements the association of the TeV emission with electrons accelerated at the termination shock of the star cluster wind of \wld \cite{HESS_Wd1_2022, Haerer2023}. 
We propose that the gamma rays in \outflow originate from electrons accelerated at the wind termination shock which accumulate in the nascent outflow.
In this scenario, the \wld superbubble expands asymmetrically due to the density gradient normal to the plane of the disk, as illustrated in Fig.~\ref{fig:sketch}.
The shock-heated material in the bubble's interior thus preferentially flows away from the Galactic mid-plane, entraining the large-scale magnetic fields.
Consequently, the accelerated particles will follow the bulk flow, subject to advection and diffusive transport.

%% discussion of gamera models
A leptonic inverse-Compton model for the spectral energy distribution of the entire region is presented, extending the model of \cite{Haerer2023} to GeV energies.
We refer to this as the \textit{total model}.
Figure~\ref{fig:sed} shows the model curve for an electron injection index of 2.25, a cluster age of 4\,Myr, wind power of $10^{39}\,\mathrm{erg}\,\mathrm{s}^{-1}$ and a constant magnetic field of 2\,$\mu$G.
Note that due to the illustrative nature of the model, these values are merely representative of the average behaviour and might be revised in future, more detailed modelling.
The required efficiency for conversion of the wind's kinetic energy to accelerated electrons at the cluster's wind termination shock is 0.7\% above 0.01\,GeV, which lies in the typical expected range.
In constructing the model, the software package \texttt{GAMERA} \cite{Hahn2015,Hahn2020} is used to model the evolution of a continuously injected particle spectrum, and the resulting gamma-ray emission.
We take into account inverse-Compton scattering on the CMB, diffuse Galactic radiation fields, and starlight from \wld (see Methods).
The contribution of synchrotron radiation to the spectrum in the energy range shown in Fig.~\ref{fig:sed} is negligible.
In the radio domain, the level of synchrotron emission predicted by the model is well below upper limits derived from radio continuum maps (see Supplementary Figure 4).

To further investigate the scenario of CR transport along the nascent outflow, we present a second model to fit only the emission of the ``far" outflow region as defined in Fig.~\ref{fig:fermi_ts_map} (\textit{far outflow model}).
We assume the same energy injection as that of the \textit{total model}, but fix the period during which injection occurred to match the normalisation in the far region.
The required time-scale of injection is 0.5--1\,Myr.
The high-energy cut-off in the far outflow spectrum can be reproduced if no new particles were injected into this region in the last approximately $125\mbox{--}200$\,kyr, corresponding to the cooling time of $\lesssim$ TeV energy electrons (see Methods).
This can be interpreted as the transport time-scale from the acceleration site (the cluster wind termination shock of \wld) to the far outflow.

The time-scale for transport to the far outflow is much shorter than expected for advection in a spherically symmetric superbubble, indicating a substantial deviation of the flow profile from $\propto 1/R^2$, or severe impact of diffusion.
Diffusion time-scales for various diffusion coefficients are given in the Methods section.
The observations presented here call for detailed multi-dimensional modelling to connect between different regions and to better constrain the transport and confinement of particles within the nascent outflow.
Ultimately, this will elucidate transport properties of CRs in superbubble environments and clarify the role of star clusters in the Galactic CR ecosystem.
Using the required injection efficiency of $0.7\%$, one can estimate the total energy density of relativistic electrons in the nascent outflow to be $U_{\rm e} \sim 1-10$ eV cm$^{-3}$.
Diffusive shock acceleration theory predicts that relativistic protons carry a larger fraction of the total energy flux, and though not measurable in the low density cavity, would carry a substantially larger energy density.
This equates to an energy density orders of magnitude greater than the locally measured CR spectrum.
This highlights the possibility that CRs can be dynamically important in regions shaped by stellar feedback.

\subsubsection*{Concluding remarks}
The newly discovered gamma-ray emission, connecting the most massive young star cluster in the Galaxy to a nascent outflow, represents a major step forward for our understanding of the role of young star clusters in CR acceleration, and of CR transport.

The new \fmlat source \outflow is consistent with the young superbubble associated to \wld being in the early stages of breaking through the edge of the Galactic Plane, and is transporting a significant population of CRs out of the disk.
The estimated CR energy density is more than an order of magnitude beyond that of the general ISM; thus CRs might dynamically influence the nascent outflow.
The observed situation is dramatically different from the typical assumption of isotropic diffusion of CRs from sources inside the disc.
It can be expected that the nascent outflow will eventually provide a channel for CRs accelerated near \wld to be transported outside of the disk into the Galactic halo.

Deeper multi-wavelength observations of the \wld system are needed to understand the evolution of the nascent outflow and hence the transport of particles into the halo.
Structures of this kind may be a common feature of massive star clusters, and gamma-ray (and multi-wavelength) searches around other systems are strongly motivated. 

\clearpage

\subsection*{Methods}

\subsubsection*{\fmlat data analysis}

The analysis shown in this paper uses 15 years of \fmlat \cite{FermiLAT2009} data from 2008 August 4 (MJD 54682) to 2023 August 3 (MJD 60159).
Time intervals during which the satellite passed through the South Atlantic Anomaly are excluded.
Our data are also filtered removing time intervals around solar flares and bright GRBs, following the procedure used in all \fmlat catalogues.
The current version of the LAT data is \texttt{P8R3}.
The event selection is based on the low-background \texttt{SOURCEVETO} class with the corresponding instrument response functions (IRFs) \texttt{P8R3\_SOURCEVETO\_V3}.
To reduce contamination from the bright gamma-ray emission from the Earth atmosphere, we select events with zenith angles smaller than $105^{\circ}$.
The data reduction and exposure calculations are performed using \emph{fermitools} version 2.2.0 and \emph{fermipy} version 1.2.0.
We perform a binned likelihood analysis combining all event types between 3\,GeV and 3\,TeV with 10 energy bins per decade.
We analyse a region of $15^{\circ} \times 15^\circ$, centred on \wld, with spatial bins of $0.03^{\circ}$, including all sources from the \fmlat 14-year Source Catalog (4FGL-DR4; \url{https://fermi.gsfc.nasa.gov/ssc/data/access/lat/14yr_catalog}) \cite{FermiLAT_4FGL_2020,FermiLAT_4FGLDR3_2022,FermiLAT_4FGLDR4_2023} in a region of $25^{\circ} \times 25^\circ$.
% We account for the effect of energy dispersion (the reconstructed energy being different from the true energy due to finite detector resolution) by setting the parameter \texttt{edisp\_bins}\,$=-1$.
% With this setting, the energy dispersion correction operates on the spectra with one extra bin below and above the threshold of the analysis\footnote{The energy dispersion correction is applied to all sources in the model, except for the isotropic diffuse emission model. More details can be found in the FSSC: \url{https://fermi.gsfc.nasa.gov/ssc/data/analysis/documentation/Pass8_edisp_usage.html}}.

Interstellar emission contributes substantially to LAT observations in the Galactic Plane, where \wld is located.
For our study of this complex region, we modelled the spatial and spectral distributions of the interstellar radiation with a linear combination of galactocentric templates (``rings'') representing the dark neutral medium (DNM), \hone, CO (i.e., gamma-ray emission from hadronic cascades produced in interactions of CRs with DNM, atomic and molecular hydrogen gas), and inverse-Compton components of the emission.
%\footnote{More details on these galactocentric templates can be found here: \url{https://fermi.gsfc.nasa.gov/ssc/data/analysis/software/aux/4fgl/Galactic_Diffuse_Emission_Model_for_the_4FGL_Catalog_Analysis.pdf}}.
These templates are those being used to derive the standard Galactic diffuse background model (\texttt{gll\_iem\_v07.fits}), but by employing separate scaling factors for the different \hone and CO rings (while using an overall normalisation for the inverse-Compton components as well as one normalisation for the positive component of the DNM), we allowed for many more degrees of freedom in fitting the diffuse emission in our region.
%\footnote{The free parameters chosen to obtain a good fit to the data after inspection of the residual maps are the following: for the CO and \hone components, we leave free both normalisation and shape for the closest rings (4, 5, 6), only the normalisation for some others (2, 3, 7, 8, 9), and completely removed rings 0 and 1 which do not contribute for this line of sight. We merged all rings for the IC component and leave free its normalisation and shape.}.
To properly extract the gamma-ray flux from the extended sources in our region, the patch component (based on the residual gamma-ray intensity from fitting the overall interstellar emission model) was not included.
The residual instrumental background and extragalactic radiation are described by a single isotropic component with spectral shape taken from the tabulated model \texttt{iso\_P8R3\_SOURCEVETO\_V3\_v1.txt}, as is standard.
% This model is available from the \emph{Fermi} Science Support Center (FSSC)\footnote{\url{https://fermi.gsfc.nasa.gov/ssc/data/access/lat/BackgroundModels.html}}.
For further details on the \fmlat data analysis, we refer to the Supplementary Material.

Our analysis starts from the baseline model provided by the 4FGL-DR4 catalogue.
The first steps of the analysis consist of a re-optimisation of the model in which we re-fit the parameters of the background models and of the brightest sources in the region of interest (ROI).
In this procedure, we also test for the presence of additional sources that are not included in the catalogue model but statistically significant in our analysis.
This is done by means of a likelihood ratio test, where the test statistic (TS) is defined as ${\rm TS}=2(\ln \mathcal{L}_1 - \ln \mathcal{L}_0)$, with $\mathcal{L}_0$ and $\mathcal{L}_1$ the likelihoods of the null hypothesis (background only) and the hypothesis being tested (source plus background), respectively.
We iteratively add 6 point-like sources with TS$>$25 to the ROI model; all are located more than two degrees away from \wld.
Their positions and TS values are included in Supplementary Table 1.

We then perform the morphological analysis of the gamma-ray emission coincident with \wld.
Since we cannot use the likelihood ratio test to compare models that are not nested, we use the Akaike Information Criterion (AIC; \cite{Akaike1974}).
We calculate $\Delta$AIC = AIC$_{1}$ - AIC$_i$ = $2 \times (\Delta$ d.o.f - $\Delta \rm{\ln \mathcal{L}})$ to compare the different models with respect to our baseline model (Model~1 in Table~\ref{tab:aic}).
Since the preferred model is the one that minimises AIC, here a larger value of $\Delta$AIC corresponds to a more favourable model.

\begin{figure}
  \centering
  \includegraphics[width=0.7\textwidth]{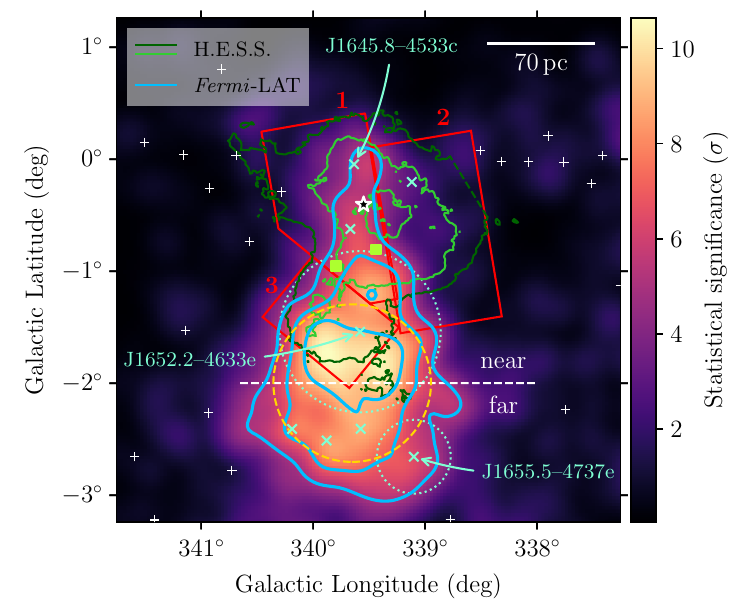}
  \caption{
    \textbf{Expanded \fmlat TS map.}
    Same as Fig.~\ref{fig:fermi_ts_map}, with additional information:
    White plus markers indicate positions of 4FGL-DR4 sources that are part of the fitted ROI model; blue-green $\times$ markers of those removed from the model.
    The dotted circles denote the extension of the disk sources 4FGL~J1652.2$-$4633e and 4FGL~J1655.5$-$4737e.
    The two green square markers show the position of 4FGL~J1648.4$-$4611 and 4FGL~J1650.3$-$4600 (associated with the pulsars PSR~J1648$-$4611 and PSR~J1650$-$4601, respectively).
    Shown in red are regions 1, 2, 3, for which separate energy spectra have been derived from \fmlat and \hess data (cf.\ Fig.~\ref{fig:sed_suppl}).
  }
  \label{fig:fermi_ts_map_suppl}
\end{figure}

We show in Fig.~\ref{fig:fermi_ts_map_suppl} a map with the positions of all 4FGL-DR4 sources in the region indicated (note that, as we will explain below, the sources marked with blue-green $\times$ markers are not part of our final model).
Several sources are located within the extent of the TeV emission detected with \hess: the two pulsars PSR~J1648$-$4611 and PSR~J1650$-$4601 (associated with 4FGL J1648.4$-$4611 and 4FGL J1650.3$-$4600, respectively), three point sources 4FGL J1644.5$-$4602c, 4FGL J1645.8$-$4533c, 4FGL J1648.4$-$4554, and the unidentified extended source 4FGL J1652.2$-$4633e.

\begin{table*}
  \caption{
    \textbf{Overview of models fitted to the \fmlat data.}
    Results of the fit of the \fmlat data between 3\,GeV and 3\,TeV using different spatial models.
    `\wld ring' refers to the region surrounding the star cluster in which TeV emission with a ring-like structure has been observed with \hess; `Outflow' denotes the putative outflow region.
    `PS' is for point source.
    The fourth column reports the log-likelihood (LLH) values obtained for each spatial model, while column~5 indicates the number of degrees of freedom adjusted in the model.
    The delta Akaike criterion, defined as $\Delta$AIC = AIC$_{1}$ - AIC$_i$ = $2 \times (\Delta$ d.o.f - $\Delta \rm{\ln \mathcal{L}})$, is reported in the last column.
  }
  \label{tab:aic} 
  \centering
  \begin{tabular}{lllccc}
    \hline \hline
    ID & \wld ring & Outflow & LLH & d.o.f & $\Delta$AIC \\
    \hline
    (1) & 3 PS & 2 disks + 3 PS & $-$840230.4 & 35 & 0 \\
    (2) & 3 PS & 1 Gauss & $-$840183.3  & 18 & 128.2 \\
    (3) & \hess temp.\ + 3 PS & 1 Gauss & $-$840120.5 & 20 & 249.8 \\
    (4) & \hess temp.\ + 1 PS & 1 Gauss & $-$840121.8 & 12 & 263.2 \\
    (5) & \hess temp.\ & 1 Gauss & $-$840128.5 & 8 &  257.8 \\
    (6) & \hess temp.\ (split) & 1 Gauss & $-$840103.8 & 10 & 303.2 \\
    (7) & \hess temp.\ (split) + 1 PS & 1 Gauss & $-$840103.7 & 14 & 295.4 \\
    (8) & \hess temp.\ (split) & 1 Gauss (split) & $-$840098.2 & 13 &  310.6 \\
    (9) & 1 disk & 1 Gauss & $-$840115.1 & 11 &  278.6 \\
    (10) & 1 disk (split) & 1 Gauss & $-$840111.3 & 13 &  282.2 \\
  \end{tabular}
\end{table*}

\begin{figure}
  \centering
  \includegraphics[width=0.85\textwidth]{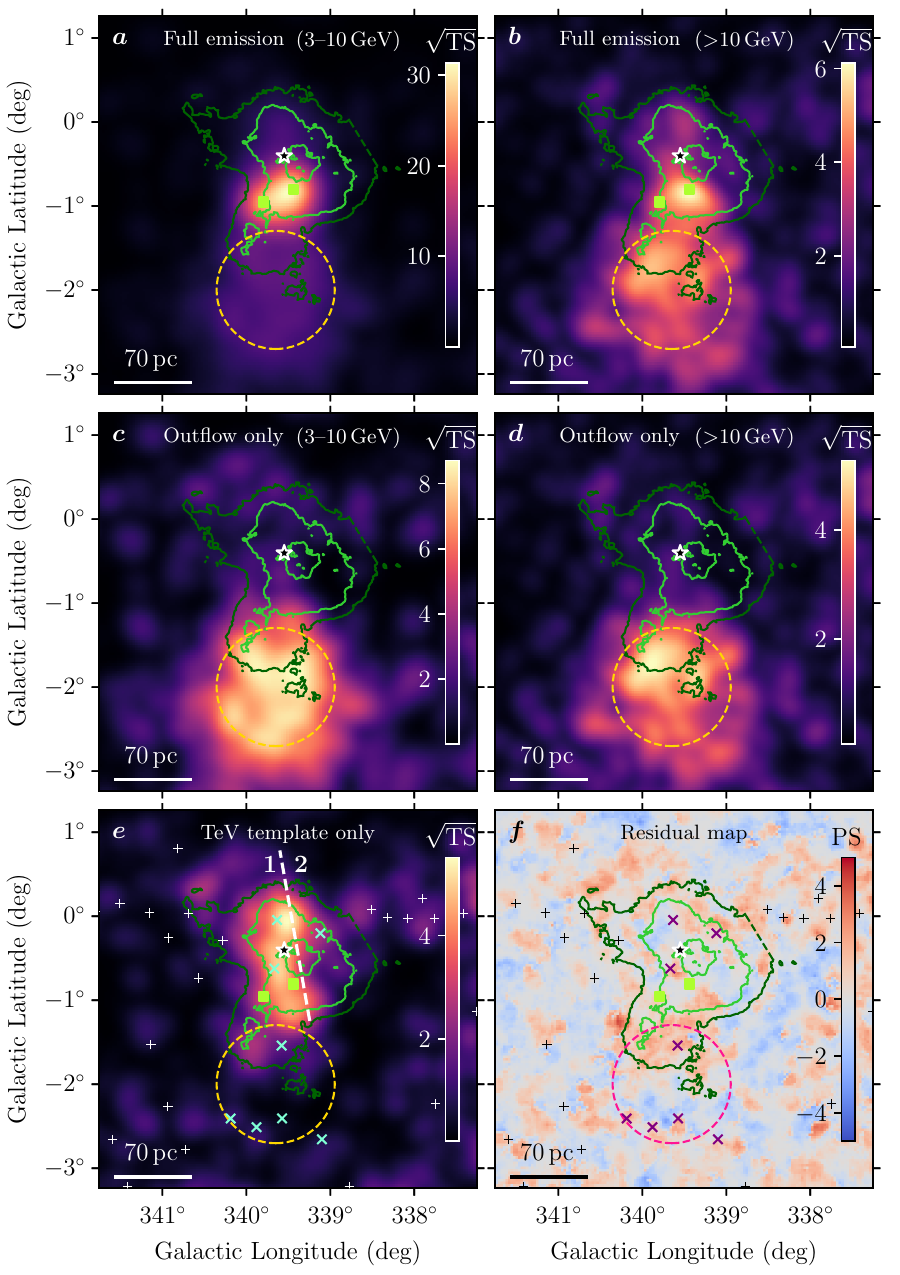}
  \caption{
    \textbf{Additional maps from the \fmlat data analysis.}
    Panels a--e show TS maps (as in Fig.~\ref{fig:fermi_ts_map_suppl}), highlighting different parts of the emission.
    (a,b): full emission, including that from the two pulsar-associated sources 4FGL~J1648.4$-$4611 and 4FGL~J1650.3$-$4600 (green square markers), in the energy range 3--10\,GeV (a) and $>$10\,GeV (b).
    (c,d): emission associated with \outflow, in the energy range 3--10\,GeV (c) and $>$10\,GeV (d).
    (e): emission surrounding \wld, associated with the template derived from the TeV emission measured with \hess.
    The white dashed line separates the template into two regions for which separate spectra are derived.
    (f): PS map \cite{Bruel2021} for the best-fit model including all components.
    Contour lines and source markers in all panels are the same as those described in Figs.~\ref{fig:fermi_ts_map},~\ref{fig:fermi_ts_map_suppl}.
  }
  \label{fig:fermi_extra_maps}
\end{figure}

As can be seen from Fig.~\ref{fig:fermi_extra_maps}(a,b), the two pulsar-associated sources 4FGL~J1648.4$-$4611 and 4FGL~J1650.3$-$4600 dominate the emission below 10\,GeV; we therefore include them in all models described in what follows.
We note, however, that the two associated pulsars have spin-down luminosities of 2--3$\times 10^{35}$\,erg\,s$^{-1}$ \cite{Manchester2005} and thus cannot be invoked to explain the extended emission present to their south, which requires a population of electrons with total power in excess of $10^{36}$\,erg\,s$^{-1}$ (see ``Modelling details'').
The centre of the extended source 4FGL~J1652.2$-$4633e lies outside the ring-like structure observed at TeV energies, but with its disk radius of $0.72^\circ$ \cite{FermiLAT_FGES_2017} the source extends up to the star cluster.
With a TS value above 3\,GeV of 770, it is very bright and requires a careful analysis.
We therefore start the morphological analysis by investigating 4FGL~J1652.2$-$4633e, keeping a log-parabola model for the spectral shape.
Interestingly, its southern edge is coincident with three unassociated point sources flagged for confusion effects (see \cite{FermiLAT_4FGL_2020}), 4FGL J1656.1$-$4706c, 4FGL J1657.7$-$4656c, and 4FGL J1658.3$-$4637c, and another unidentified extended source 4FGL J1655.5$-$4737e represented by a radial disk (cf.\ Fig.~\ref{fig:fermi_ts_map_suppl}).
By replacing these 5 model components with a simple Gaussian (Model~2 in Table~\ref{tab:aic}), the likelihood of the fit significantly improves while the number of degrees of freedom is reduced with respect to the baseline model (the large number of degrees of freedom for Model~1 comes about because we re-fit the positions of all sources).
This Gaussian component is referred to with \outflow in this work.
Its best-fit Galactic longitude and latitude are $l=(339.61 \pm  0.03)^\circ$ and $b=(-1.91 \pm 0.04)^\circ$, respectively, while the best-fit 1-$\sigma$ radius is $(0.71 \pm 0.03)^\circ$.
We note that diffuse gamma-ray emission with a somewhat smaller extent, coincident with the TeV emission protruding from the ``ring'' measured with \hess, had already been reported in \cite{Ohm2013}.

Moving to the vicinity of the cluster, the next step is to use the \hess flux map \cite{HESS_Wd1_2022} as a template to fit the \fmlat data, along with the three unidentified point sources located within the TeV ring structure (Model~3).
We find that keeping only one of the three sources overlapping with the \hess template, coincident with 4FGL J1645.8$-$4533c, is sufficient (but required) to provide smooth residuals (Model~4/Model~5).
Fig.~\ref{fig:fermi_extra_maps}(e) illustrates that the GeV emission surrounding the cluster is generally compatible in spatial structure with the TeV emission detected by \hess, but brighter in region~1 (i.e. left of the dashed white line) than in region~2.
We therefore split the \hess template along the division between regions~1 and~2 and fit them separately, thus providing an improved likelihood value (Model~6).
We verify that with a split template, there is no need for the additional point source coincident with 4FGL J1645.8$-$4533c any more (Model~7).
To test the scenario in which \outflow is produced by energy-dependent cooling of electrons, we also divide the Gaussian model in two parts, parallel to the Galactic Plane (along the near/far division indicated in Fig.~\ref{fig:fermi_ts_map_suppl}), providing our best model of the region (Model~8) following the Akaike criterium.
Model 8 provides a $2.5\sigma$ improvement ($\Delta$TS of 11.2 for 3 additional degrees of freedom) with respect to Model 6.
The PS map (see Fig.~\ref{fig:fermi_extra_maps}(f)), constructed to display both positive and negative residuals with respect to the final model \cite{Bruel2021}, demonstrates that this model leads to no statistically significant residuals with respect to the observed data.
% \footnote{Similar to the TS map, the PS map shows residual emission with respect to the model prediction. In contrast to the TS map, however, the PS map also displays negative residuals. The PS value can be converted to a statistical significance. For more details, see \cite{Bruel2021}.} shown in Fig.~\ref{fig:additional_fermi_maps}(b) demonstrates that this model leads to no statistically significant residuals with respect to the observed data.
Dividing the Gaussian source into smaller slices does not provide any significant improvement in the fit.
Finally, as an additional cross-check we use a simple disk model to fit the gamma-ray emission associated with \wld (Model~9), as well as a disk divided into 2 regions (Model~10). While the disk model yields a reasonable fit, the best model remains the \hess template divided in 2 parts (Model 8).

We test a simple power-law model and a power law with an exponential cut-off for each component of our best spatial model (i.e., the two halves of the \hess template for the region around \wld and the two halves of the Gaussian model for the outflow region).
The results from these fits are provided in Supplementary Table 2, and corresponding SEDs including systematic uncertainties are displayed in Supplementary Figure 2.

Concerning \outflow, we find that the spectrum in the near part extends to higher energies, while there is an indication of a cut-off to the spectrum in the far part.
This is in line with our observation that the spatial modelling is improved when splitting \outflow into a near and far part.
The maps in Fig.~\ref{fig:fermi_extra_maps}(c,d), which show the spatial morphology of \outflow in two energy ranges, further support this: above 10\,GeV, the far part is less bright than the near one, which is not the case at lower energies.
A statistical test yields a significance of around 2$\sigma$ for this difference.
Thus, we find an indication of a spectral softening of the emission along the putative outflow, although we cannot claim this with high significance.

The apparent asymmetry in the spectra derived for \wld in regions 1 and 2 could originate from differences in how electrons are injected in both regions.
That region~2 is equally bright as region~1 in the TeV domain but dimmer in the GeV domain would then imply that only freshly injected electrons are visible in this region.
Alternatively, a deviation in spectral index was detected by \cite{HESS_Wd1_2022} in a region close to LMXB 4U 1642$-$45, which is located in region~1 (see also the recent investigation of this region with eROSITA \cite{Haubner2025}).
The region also coincides with the soft and confused 4FGL source J1645.8$-$4533c.
The current statistics does not allow the detection of an additional point source within region~1 (see Table~\ref{tab:aic}), but an unidentified, soft-spectrum source could contaminate the spectrum of \wld (region 1) at low energy, thus explaining the asymmetry between regions~1 and~2.

\clearpage

%===================================================%
\subsubsection*{Comparison between \fmlat and \hess SEDs}

\begin{figure}[bh]
  \centering
  \includegraphics[width=0.7\linewidth]{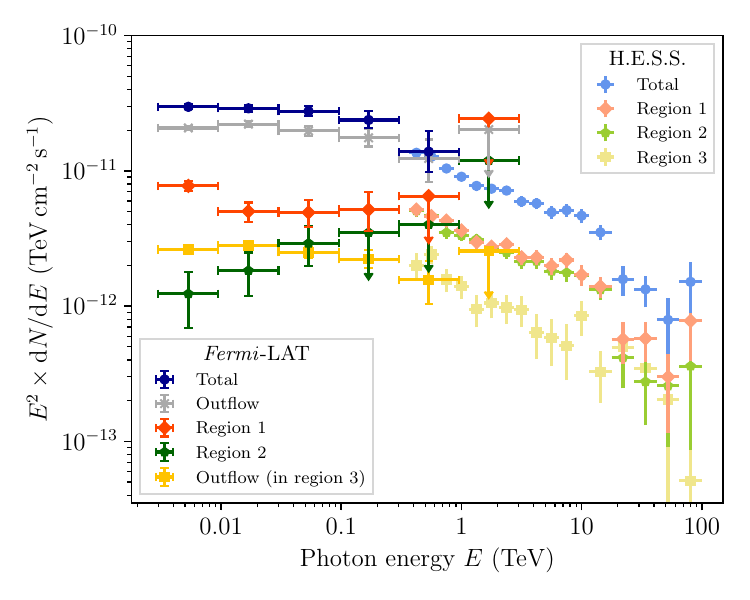}
  \caption{
    \textbf{Comparison of energy spectra derived from \fmlat and \hess data.}
    Region labels refer to regions as shown in Fig.~\ref{fig:fermi_ts_map_suppl}.
    Error bars denote 68\% c.l.\ statistical uncertainties; upper limits are at 95\% c.l.
  }
  \label{fig:sed_suppl}
\end{figure}

In addition to the comparison of the total gamma-ray emission measured with \fmlat and \hess shown in Fig.~\ref{fig:sed}, we have also compared spectra for different spatial regions (labelled 1, 2, 3 in Fig.~\ref{fig:fermi_ts_map_suppl}).
This comparison is shown in Fig.~\ref{fig:sed_suppl}, which demonstrates that the spectra in all regions connect smoothly between the energy ranges covered by the two instruments.

\fmlat spectra for regions~1 and~2 are a direct result of the modelling described in the previous subsection, where Models 6--8 utilise a \hess flux map template divided along the separating line between the two regions.
The SED for region~3 has been computed by integrating the Gaussian spatial model in this region and scaling the total SED by the ratio of that integral to the full model.

To derive the \hess SEDs, we sum up the spectra of the square regions a--p defined in \cite{HESS_Wd1_2022}, taking into account the overlap between the square regions and the regions 1--3 defined here.
This assumes that the gamma-ray flux is isotropic within each square region, which is the case in good approximation.

\clearpage

%==========================================================%
\subsubsection*{Galactic kinematics in the direction of Westerlund~1}

In order to study the correlation of gamma-ray emission in the \outflow region with ISM structures we use emission lines from \hone and CO.
Based on the plausible physical connection between \outflow and \wld suggested by the gamma-ray morphology and our modelling work, we aim at studying the ISM in the vicinity of \wld.
Therefore, we need to identify the range in Doppler-shift velocity with respect to the local standard of rest ($v_\mathrm{LSR}$) of the lines corresponding to the region of \wld.
It is well known that the kinematics of gas and stars in the direction of \wld do not follow closely large-scale rotation curves \cite{Reid2019,Negueruela2022}.
If we convert the measured distance of $d=4.14$\,kpc (i.e.\ the average of the estimates presented in \cite{Negueruela2022,Navarete2022}) to a velocity, we obtain $v_\mathrm{LSR}\approx -61.7$\,km\,s$^{-1}$, where we have used the rotation curve model from \cite{RomanDuval2009} with an orbital velocity of the Sun $V_0=220$\,km\,s$^{-1}$ and a galactocentric distance of the Sun $R_0=8.178\,\mathrm{kpc}$ \cite{GRAVITY2019}.
This is inconsistent with the velocity measured from stars in the cluster itself, which is around $v_\mathrm{LSR}\approx-43$\,km\,s$^{-1}$ \cite{Negueruela2022}.
Other kinematic tracers like IR masers and \hii regions potentially associated with \wld also display velocities in the range from $v_\mathrm{LSR}\approx-50$\,km\,s$^{-1}$ to $v_\mathrm{LSR}\approx-38$\,km\,s$^{-1}$ \cite{Russeil2003,Fok2012}.

A dedicated analysis of \hone properties in the environment of \wld \cite{Kothes2007} led to the identification of two bubble-like features, dubbed B1 and B2, spatially associated with \wld at radial velocities $\approx -55$\,km\,s$^{-1}$, that is $v_\mathrm{LSR}\approx -51$\,km\,s$^{-1}$, and expanding with a velocity of $\approx 5$\,km\,s$^{-1}$.
These bubble-like features are interpreted as cavities in the ISM carved by activity of \wld itself.
In the same velocity range, a third bubble, B3, extends to lower Galactic latitudes.
To the north, B3 consists of a shell with two large, bright, and complex emission regions to the east and west, while it is open to the south, away from the Galactic Plane \cite{Kothes2007} in the direction of \outflow.
The same study shows the complexity of \hone kinematic structures in this region severely affected by the near-far ambiguity.
In the kinematic range discussed above \hone emission appears to arise from the superposition of the Norma and Scutum-Crux arms, with \wld probably located on the near side of the Norma arm based on recent studies of spiral tracers \cite{Hou2021} and the distance estimate from cluster member parallaxes \cite{Negueruela2022}. 

\begin{figure}[bh]
  \centering
  \includegraphics[width=0.7\linewidth]{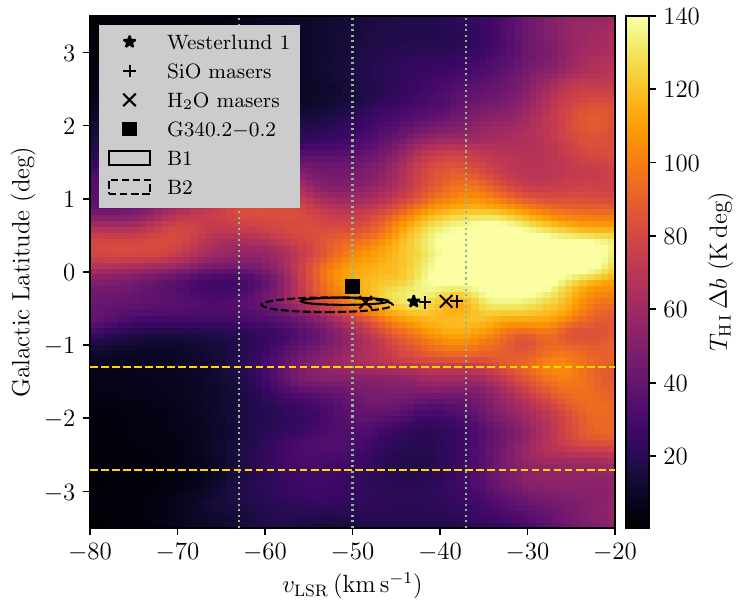}
  \caption{
    \textbf{Velocity-latitude diagram of \hone emission in the region around \wld and \outflow.}
    The \hone brightness temperatures \cite{Kalberla2015} are integrated over the $1\sigma$ longitude range of \outflow.
    We show the position of \wld member stars \cite{Negueruela2022}, SiO and H$_2$O masers W~26 and W~237 \cite{Fok2012}, the \hii region G340.2$-$0.2 \cite{Russeil2003}, and the \hone bubble-like features B1 and B2 \cite{Kothes2007}.
    Dashed horizontal lines show the $1\sigma$ latitude range of \outflow.
    Dotted vertical lines show the boundaries of the velocity integration ranges used for column-density maps in our study.  
  }
  \label{fig:vbdiag}
\end{figure}

In this work we use \hone data from the Parkes Galactic All-Sky Survey third data release (GASS~III) \cite{Kalberla2015} as other higher-resolution surveys do not sufficiently cover the region of \outflow.
Figure~\ref{fig:vbdiag} shows a velocity-latitude diagram of \hone emission in a large region around \wld and \outflow.
For the entire velocity range spanned by kinematic tracers associated with \wld and by the bubble-like features B1 and B2 from \cite{Kothes2007} the $1\sigma$ latitude range of \outflow corresponds to low-emission regions just below the bright emission band from the Galactic Plane.

In acknowledgement of the substantial uncertainties related to the identification of gas in the vicinity of \wld and \outflow we present gas maps throughout the paper in two velocity ranges, namely $v_\mathrm{LSR} = [-63, -50]$\,km\,s$^{-1}$ and $v_\mathrm{LSR} = [-50, -37]$\,km\,s$^{-1}$.
These ranges are also indicated in Fig.~\ref{fig:vbdiag}.

\clearpage

%===============================================%
\subsubsection*{H {\sc i} column densities in the J1654--467 region}

Employing again the GASS~III data, we provide in Fig.~\ref{fig:gas_maps} maps displaying the column density of atomic hydrogen in our region of interest.
To convert from brightness temperature to column density, we assume that the gas is optically thin and employ a conversion factor of $X_\mathrm{HI}=1.823\times 10^{18}\,\mathrm{cm}^{-2}\,/\,(\mathrm{K}\,\mathrm{km}\,\mathrm{s}^{-1})$.
We verified that, at the high latitude and modest total column density at the location of \outflow, qualitatively and quantitatively compatible maps are obtained when using a low spin temperature of 100\,K.

An under-density with respect to neighbouring lines of sight is visible at the location of \outflow in both velocity ranges displayed in Fig.~\ref{fig:gas_maps}.
In particular, a clear minimum in density that coincides with the peak of GeV gamma-ray emission can be identified in the left panel.
In order to quantify this observation, we compute a rough estimate of the difference in density between the outflow region and its surroundings.
We take the region indicated by the yellow dashed circle in Fig.~\ref{fig:gas_maps} (which corresponds to the 1-$\sigma$ radius of the Gaussian model used to describe \outflow) as representative for the outflow and place two control regions (shown by the white lines) that together cover the same solid angle.
For the first velocity interval ($[-63,-50]$\,km\,s$^{-1}$, left panel), the average column density in the outflow region is $\approx 5.2\times 10^{20}\,\mathrm{cm}^{-2}$, while it is $\approx 6.7\times 10^{20}\,\mathrm{cm}^{-2}$ in the control regions.
Hence, the outflow region is less dense by about $1.5\times 10^{20}\,\mathrm{cm}^{-2}$.
Assuming an extent of the cavity along the line of sight of 70\,pc, this would correspond to about $0.7$~atoms\,cm$^{-3}$ excavated by the outflow.
Performing the same calculation for the second velocity interval ($[-50,-37]$\,km\,s$^{-1}$, right panel), we obtain a difference in density of $\approx 0.7\times 10^{20}\,\mathrm{cm}^{-2}$, or about 0.3\,atoms\,cm$^{-3}$ excavated.
Thus, we conclude that the outflow region is less dense by about 0.3--0.7\,atoms\,cm$^{-3}$ compared to its surroundings.
We note that the larger difference in the first velocity interval –– which corresponds to larger distances from the Earth -- may indicate that the tentative outflow is not directed completely perpendicular to the line of sight, but exhibits a component directed away from us.

\clearpage

%============================%
\subsubsection*{Modelling details}

We use the software package \texttt{GAMERA} \cite{Hahn2015,Hahn2020} to compute the time evolution of a continuously injected particle spectrum and compute the resulting gamma-ray emission.
The total and far outflow spectra (see Fig.~\ref{fig:sed}) are described by two independent one-zone models.
We consider inverse-Compton scattering on the CMB ($T=2.7\,$K, $U_{\rm CMB}=0.26\,\mathrm{eV}\,\mathrm{cm}^{-3}$) and diffuse starlight and dust-scattered starlight \cite{Popescu2017}.
The electron injection spectrum is $\mathrm{d}N/(\mathrm{d}E\mathrm{d}t) =\eta \left(E /E_0\right)^{-2.25} \exp(-E/170\,\mathrm{TeV})$, where the cut-off is set by the balance of acceleration and loss timescale and $\eta$ is an efficiency parameter corresponding to the fraction of star cluster wind power transferred to electrons in the acceleration process, $L_\mathrm{inj} = \eta L_\mathrm{w}$.
We use $\eta = 0.7\,\%$ for $L_\mathrm{w} = 10^{39}\,\mathrm{erg}\,\mathrm{s}^{-1}$ above $0.01\,$GeV for the model shown in Fig.~\ref{fig:sed}. 
For further details on the modelling, justification for the choice of parameters, and a discussion of the hadronic scenario see \citet{Haerer2023}. 

\begin{figure}[bh]
  \centering
  \includegraphics[width=0.9\linewidth]{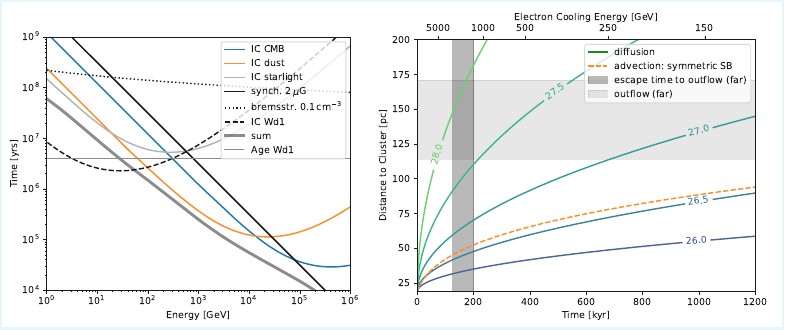}
  \caption{
    \textbf{Cooling and transport timescales.}
    Left: Cooling timescales.
    Inverse-Compton emission on cluster photon field (IC Wd1) is evaluated at the wind termination shock (20\,pc).
    The sum only includes components that are relevant at the location of \outflow.
    Right: Transport timescales.
    The coloured solid lines indicate the transport for a range of diffusion coefficients, $D$, and are labelled with $\log_{10} \left( D/\mathrm{cm}^2\,\mathrm{s}^{-1} \right)$.
    The orange dashed line indicates advection in a spherically symmetric superbubble (SB).
    The upper abscissa shows the energy of electrons that cool within the time given on the lower abscissa.
    For further details see the text.
  }
  \label{fig:scales}
\end{figure}

The left-hand plot of Fig.~\ref{fig:scales} shows the timescales for competing electron cooling mechanisms. 
The photon field of the massive stars in \wld is assumed to have an effective temperature of $T_\mathrm{eff}=40{\,}000\,$K.
In the \textit{total model}, the cluster photon field is evaluated at the cluster wind termination shock (20\,pc), where the energy density is $U_{\rm ph}=42\,\mathrm{eV}\,\mathrm{cm}^{-3}$.
Since $U_{\rm ph}\propto R^{-2}$, the cluster photon field is negligible in the nascent outflow and is therefore also negligible in the \textit{far region model}.
The right-hand plot of Fig.~\ref{fig:scales} shows transport timescales for diffusion and advection inside the \wld superbubble.
We apply the same assumptions as in \citet{Haerer2023}, for both diffusion and advection.
Advection adopts a cluster wind velocity of 2500$\,\mathrm{km}\,\mathrm{s}^{-1}$ and a scaling of the flow speed $\propto R^{-2}$ beyond the cluster wind termination shock.
We plot also the transport times for diffusion coefficients $D=10^\alpha\,\mathrm{cm^2}\,\mathrm{s}^{-1}$ with exponents in the range $\alpha=26\mbox{--}28$.
Adopting Kolmogorov scaling in a fully turbulent field (i.e. no strong guide field), the electron diffusion coefficient is
\begin{equation}
    D = 2.5\times 10^{26} \left(\frac{E}{1\,\mathrm{GeV}} \right)^{1/3} \left( \frac{B}{2\,\mu\mathrm{G}} \right)^{-1/3} \left( \frac{R_\mathrm{inj}}{1\,\mathrm{pc}} \right)^{2/3} \, \mathrm{cm}^2 \, \mathrm{s}^{-1} \, ,
\end{equation}
where $B$ is the average magnetic field strength and $R_\mathrm{inj}$ the turbulence injection scale.
$D$ takes values of $1.2\times10^{27} \, \mathrm{cm}^2 \, \mathrm{s}^{-1}$ at 100\,GeV and $2.5\times10^{27} \, \mathrm{cm}^2 \, \mathrm{s}^{-1}$ at 1\,TeV.
Applying the same approach using Kraichnan scaling suggests diffusion coefficients of $2.3\times10^{26} \, \mathrm{cm}^2 \, \mathrm{s}^{-1}$ at 100\,GeV and $7.2\times10^{26} \, \mathrm{cm}^2 \, \mathrm{s}^{-1}$ at 1\,TeV.
The dark grey band in Fig.~\ref{fig:scales} marks the expected escape time to the far outflow, obtained from the \textit{far region model} as described in the main text.
The lower bound (125\,kyr) corresponds to the model shown in Fig.~\ref{fig:sed}.
The upper bound (200\,kyr) is obtained by changing the normalisation of the model for the entire region to match the \fmlat points above 10\,GeV.
The canonical model for advection in the superbubble (orange dashed line) under-predicts the transport length-scale (about $40\mbox{--}50\,$pc compared to ${\gtrsim}110\,$pc).

\clearpage

\backmatter

\bmhead{Data Availability}
The \fmlat data and software needed for analysis are available from the \textit{Fermi} Science Support Center, \url{https://fermi.gsfc.nasa.gov/ssc}.
All other data generated or analysed during this study are included in this published article (and its supplementary information files), or referenced in the article itself.

\bmhead{Acknowledgements}
The \fmlat Collaboration acknowledges generous ongoing support from a number of agencies and institutes that have supported both the development and the operation of the LAT as well as scientific data analysis.
These include
the National Aeronautics and Space Administration and the Department of Energy in the United States,
the Commissariat \`a l'Energie Atomique and the Centre National de la Recherche Scientifique / Institut National de Physique
Nucl\'eaire et de Physique des Particules in France,
the Agenzia Spaziale Italiana and the Istituto Nazionale di Fisica Nucleare in Italy,
the Ministry of Education, Culture, Sports, Science and Technology (MEXT), High Energy Accelerator Research
Organization (KEK) and Japan Aerospace Exploration Agency (JAXA) in Japan,
and the K.~A.~Wallenberg Foundation, the Swedish Research Council and the Swedish National Space Board in Sweden.
Additional support for science analysis during the operations phase is gratefully acknowledged from the Istituto Nazionale di Astrofisica in Italy and the Centre National d'\'Etudes Spatiales in France.
This work performed in part under DOE Contract DE-AC02-76SF00515.

Marianne Lemoine-Goumard acknowledges support from the Alexander von Humboldt Foundation.

The authors thank Jes\'us Ma\'iz Apell\'aniz, Ignacio Negueruela D\'iez, and Quentin Remy for discussions about the interstellar gas in the vicinity of \wld.

\bmhead{Author Contributions Statement}
M.L.-G.\ performed the \fmlat data analysis.
L.M.\ carried out the study of the ISM density and created all figures (except Fig.~\ref{fig:scales}).
L.H.\ developed the theoretical modelling and produced Fig.~\ref{fig:scales}, with support by B.R.\ and T.V.
R.B., G.P., and L.T.\ assisted with the \fmlat data analysis; L.T.\ furthermore helped with the ISM density study.
M.L.-G., L.H., L.M., J.H., and B.R.\ wrote the paper.
All authors have provided comments to initial versions of the article.

\bmhead{Competing Interests Statement}
The authors declare no competing interests.

\clearpage

% \section*{Declarations}

% Some journals require declarations to be submitted in a standardised format. Please check the Instructions for Authors of the journal to which you are submitting to see if you need to complete this section. If yes, your manuscript must contain the following sections under the heading `Declarations':

% \begin{itemize}
% \item Conflict of interest/Competing interests (check journal-specific guidelines for which heading to use)
% \item Ethics approval and consent to participate
% \item Consent for publication
% \item Materials availability
% \end{itemize}

% \noindent
% If any of the sections are not relevant to your manuscript, please include the heading and write `Not applicable' for that section. 


\makeatletter

% avoid page breaks in bib items
\interlinepenalty=10000

\makeatother


\begin{thebibliography}{63}
\ifx \bisbn   \undefined \def \bisbn  #1{ISBN #1}\fi
\ifx \binits  \undefined \def \binits#1{#1}\fi
\ifx \bauthor  \undefined \def \bauthor#1{#1}\fi
\ifx \batitle  \undefined \def \batitle#1{#1}\fi
\ifx \bjtitle  \undefined \def \bjtitle#1{#1}\fi
\ifx \bvolume  \undefined \def \bvolume#1{\textbf{#1}}\fi
\ifx \byear  \undefined \def \byear#1{#1}\fi
\ifx \bissue  \undefined \def \bissue#1{#1}\fi
\ifx \bfpage  \undefined \def \bfpage#1{#1}\fi
\ifx \blpage  \undefined \def \blpage #1{#1}\fi
\ifx \burl  \undefined \def \burl#1{\textsf{#1}}\fi
\ifx \doiurl  \undefined \def \doiurl#1{\href{https://doi.org/#1}{#1}}\fi
\ifx \betal  \undefined \def \betal{\textit{et al.}}\fi
\ifx \binstitute  \undefined \def \binstitute#1{#1}\fi
\ifx \binstitutionaled  \undefined \def \binstitutionaled#1{#1}\fi
\ifx \bctitle  \undefined \def \bctitle#1{#1}\fi
\ifx \beditor  \undefined \def \beditor#1{#1}\fi
\ifx \bpublisher  \undefined \def \bpublisher#1{#1}\fi
\ifx \bbtitle  \undefined \def \bbtitle#1{#1}\fi
\ifx \bedition  \undefined \def \bedition#1{#1}\fi
\ifx \bseriesno  \undefined \def \bseriesno#1{#1}\fi
\ifx \blocation  \undefined \def \blocation#1{#1}\fi
\ifx \bsertitle  \undefined \def \bsertitle#1{#1}\fi
\ifx \bsnm \undefined \def \bsnm#1{#1}\fi
\ifx \bsuffix \undefined \def \bsuffix#1{#1}\fi
\ifx \bparticle \undefined \def \bparticle#1{#1}\fi
\ifx \barticle \undefined \def \barticle#1{#1}\fi
\bibcommenthead
\ifx \bconfdate \undefined \def \bconfdate #1{#1}\fi
\ifx \botherref \undefined \def \botherref #1{#1}\fi
\ifx \url \undefined \def \url#1{\textsf{#1}}\fi
\ifx \bchapter \undefined \def \bchapter#1{#1}\fi
\ifx \bbook \undefined \def \bbook#1{#1}\fi
\ifx \bcomment \undefined \def \bcomment#1{#1}\fi
\ifx \oauthor \undefined \def \oauthor#1{#1}\fi
\ifx \citeauthoryear \undefined \def \citeauthoryear#1{#1}\fi
\ifx \endbibitem  \undefined \def \endbibitem {}\fi
\ifx \bconflocation  \undefined \def \bconflocation#1{#1}\fi
\ifx \arxivurl  \undefined \def \arxivurl#1{\textsf{#1}}\fi
\csname PreBibitemsHook\endcsname

%%% 1
\bibitem[\protect\citeauthoryear{{Thompson} and
  {Heckman}}{2024}]{ThompsonHeckman}
\begin{barticle}
\bauthor{\bsnm{{Thompson}}, \binits{T.A.}},
\bauthor{\bsnm{{Heckman}}, \binits{T.M.}}:
\batitle{{Theory and Observation of Winds from Star-Forming Galaxies}}.
\bjtitle{\araa}
\bvolume{62},
\bfpage{529}--\blpage{591}
(\byear{2024})
DOI: \doiurl{10.1146/annurev-astro-041224-011924}
{arXiv:\href{https://arxiv.org/abs/2406.08561}{{2406.08561}}}
\end{barticle}
\endbibitem

%%% 2
\bibitem[\protect\citeauthoryear{Girichidis et~al.}{2016}]{Girichidis}
\begin{barticle}
\bauthor{\bsnm{Girichidis}, \binits{P.}}, \betal:
\batitle{{Launching Cosmic-Ray-driven Outflows from the Magnetized Interstellar
  Medium}}.
\bjtitle{\apjl}
\bvolume{816},
\bfpage{19}
(\byear{2016})
DOI: \doiurl{10.3847/2041-8205/816/2/L19}
{arXiv:\href{https://arxiv.org/abs/1509.07247}{{1509.07247}}}
\end{barticle}
\endbibitem

%%% 3
\bibitem[\protect\citeauthoryear{{Rathjen} et~al.}{2021}]{Rathjen2021}
\begin{barticle}
\bauthor{\bsnm{{Rathjen}}, \binits{T.-E.}}, \betal:
\batitle{{SILCC VI - Multiphase ISM structure, stellar clustering, and outflows
  with supernovae, stellar winds, ionizing radiation, and cosmic rays}}.
\bjtitle{\mnras}
\bvolume{504},
\bfpage{1039}--\blpage{1061}
(\byear{2021})
DOI: \doiurl{10.1093/mnras/stab900}
{arXiv:\href{https://arxiv.org/abs/2103.14128}{{2103.14128}}}
\end{barticle}
\endbibitem

%%% 4
\bibitem[\protect\citeauthoryear{{Modak} et~al.}{2023}]{Modak}
\begin{barticle}
\bauthor{\bsnm{{Modak}}, \binits{S.}},
\bauthor{\bsnm{{Quataert}}, \binits{E.}},
\bauthor{\bsnm{{Jiang}}, \binits{Y.-F.}},
\bauthor{\bsnm{{Thompson}}, \binits{T.A.}}:
\batitle{{Cosmic-ray driven galactic winds from the warm interstellar medium}}.
\bjtitle{\mnras}
\bvolume{524},
\bfpage{6374}--\blpage{6391}
(\byear{2023})
DOI: \doiurl{10.1093/mnras/stad2257}
{arXiv:\href{https://arxiv.org/abs/2302.03701}{{2302.03701}}}
\end{barticle}
\endbibitem

%%% 5
\bibitem[\protect\citeauthoryear{{Armillotta} et~al.}{2024}]{Armilotta}
\begin{barticle}
\bauthor{\bsnm{{Armillotta}}, \binits{L.}},
\bauthor{\bsnm{{Ostriker}}, \binits{E.C.}},
\bauthor{\bsnm{{Kim}}, \binits{C.-G.}},
\bauthor{\bsnm{{Jiang}}, \binits{Y.-F.}}:
\batitle{{Cosmic-Ray Acceleration of Galactic Outflows in Multiphase Gas}}.
\bjtitle{\apj}
\bvolume{964},
\bfpage{99}
(\byear{2024})
DOI: \doiurl{10.3847/1538-4357/ad1e5c}
{arXiv:\href{https://arxiv.org/abs/2401.04169}{{2401.04169}}}
\end{barticle}
\endbibitem

%%% 6
\bibitem[\protect\citeauthoryear{Sike et~al.}{2024}]{Sike2024}
\begin{botherref}
\oauthor{\bsnm{Sike}, \binits{B.}},
\oauthor{\bsnm{Thomas}, \binits{T.}},
\oauthor{\bsnm{Ruszkowski}, \binits{M.}},
\oauthor{\bsnm{Pfrommer}, \binits{C.}},
\oauthor{\bsnm{Weber}, \binits{M.}}:
{Cosmic Ray-Driven Galactic Winds with Resolved ISM and Ion-Neutral Damping}.
Accepted for publication in \apj
(2024)
{arXiv:\href{https://arxiv.org/abs/2410.06988}{{2410.06988}}}
\end{botherref}
\endbibitem

%%% 7
\bibitem[\protect\citeauthoryear{Kjellgren et~al.}{2025}]{Kjellgren2025}
\begin{botherref}
\oauthor{\bsnm{Kjellgren}, \binits{K.}}, et al.:
{The dynamical impact of cosmic rays in the Rhea magnetohydrodynamics
  simulations}.
Submitted to \aap
(2025)
{arXiv:\href{https://arxiv.org/abs/2502.02635}{{2502.02635}}}
\end{botherref}
\endbibitem

%%% 8
\bibitem[\protect\citeauthoryear{{Ruszkowski} and
  {Pfrommer}}{2023}]{Ruszkowski}
\begin{barticle}
\bauthor{\bsnm{{Ruszkowski}}, \binits{M.}},
\bauthor{\bsnm{{Pfrommer}}, \binits{C.}}:
\batitle{{Cosmic ray feedback in galaxies and galaxy clusters}}.
\bjtitle{\aapr}
\bvolume{31},
\bfpage{4}
(\byear{2023})
DOI: \doiurl{10.1007/s00159-023-00149-2}
{arXiv:\href{https://arxiv.org/abs/2306.03141}{{2306.03141}}}
\end{barticle}
\endbibitem

%%% 9
\bibitem[\protect\citeauthoryear{{Johnson} and {Axford}}{1971}]{JohnsonAxford}
\begin{barticle}
\bauthor{\bsnm{{Johnson}}, \binits{H.E.}},
\bauthor{\bsnm{{Axford}}, \binits{W.I.}}:
\batitle{{Galactic Winds}}.
\bjtitle{\apj}
\bvolume{165},
\bfpage{381}
(\byear{1971})
DOI: \doiurl{10.1086/150903}
\end{barticle}
\endbibitem

%%% 10
\bibitem[\protect\citeauthoryear{{Ipavich}}{1975}]{Ipavich}
\begin{barticle}
\bauthor{\bsnm{{Ipavich}}, \binits{F.M.}}:
\batitle{{Galactic winds driven by cosmic rays}}.
\bjtitle{\apj}
\bvolume{196},
\bfpage{107}--\blpage{120}
(\byear{1975})
DOI: \doiurl{10.1086/153397}
\end{barticle}
\endbibitem

%%% 11
\bibitem[\protect\citeauthoryear{Breitschwerdt et~al.}{1991}]{BreitschwerdtI}
\begin{barticle}
\bauthor{\bsnm{Breitschwerdt}, \binits{D.}},
\bauthor{\bsnm{McKenzie}, \binits{J.F.}},
\bauthor{\bsnm{V\"olk}, \binits{H.J.}}:
\batitle{{Galactic winds. I. Cosmic ray and wave-driven winds from the
  galaxy}}.
\bjtitle{\aap}
\bvolume{245},
\bfpage{79}
(\byear{1991})
\end{barticle}
\endbibitem

%%% 12
\bibitem[\protect\citeauthoryear{Breitschwerdt et~al.}{1993}]{BreitschwerdtII}
\begin{barticle}
\bauthor{\bsnm{Breitschwerdt}, \binits{D.}},
\bauthor{\bsnm{McKenzie}, \binits{J.F.}},
\bauthor{\bsnm{V\"olk}, \binits{H.J.}}:
\batitle{{Galactic winds. II. Role of the disk-halo interface in cosmic ray
  driven galactic winds}}.
\bjtitle{\aap}
\bvolume{269},
\bfpage{54}--\blpage{66}
(\byear{1993})
\end{barticle}
\endbibitem

%%% 13
\bibitem[\protect\citeauthoryear{McKee and Ostriker}{1977}]{McKee1977}
\begin{barticle}
\bauthor{\bsnm{McKee}, \binits{C.F.}},
\bauthor{\bsnm{Ostriker}, \binits{J.P.}}:
\batitle{A theory of the interstellar medium: three components regulated by
  supernova explosions in an inhomogeneous substrate}.
\bjtitle{\apj}
\bvolume{218},
\bfpage{148}--\blpage{169}
(\byear{1977})
DOI: \doiurl{10.1086/155667}
\end{barticle}
\endbibitem

%%% 14
\bibitem[\protect\citeauthoryear{{Ferri{\`e}re}}{2001}]{Ferriere}
\begin{barticle}
\bauthor{\bsnm{{Ferri{\`e}re}}, \binits{K.M.}}:
\batitle{{The interstellar environment of our galaxy}}.
\bjtitle{Reviews of Modern Physics}
\bvolume{73},
\bfpage{1031}--\blpage{1066}
(\byear{2001})
DOI: \doiurl{10.1103/RevModPhys.73.1031}
{arXiv:\href{https://arxiv.org/abs/astro-ph/0106359}{{astro-ph/0106359}}}
\end{barticle}
\endbibitem

%%% 15
\bibitem[\protect\citeauthoryear{Mac~Low et~al.}{1989}]{MacLow1989}
\begin{barticle}
\bauthor{\bsnm{Mac~Low}, \binits{M.-M.}},
\bauthor{\bsnm{McCray}, \binits{R.}},
\bauthor{\bsnm{Norman}, \binits{M.L.}}:
\batitle{{Superbubble Blowout Dynamics}}.
\bjtitle{\apj}
\bvolume{337},
\bfpage{141}
(\byear{1989})
DOI: \doiurl{10.1086/167094}
\end{barticle}
\endbibitem

%%% 16
\bibitem[\protect\citeauthoryear{{Gaensler} et~al.}{2008}]{Gaensler08}
\begin{barticle}
\bauthor{\bsnm{{Gaensler}}, \binits{B.M.}},
\bauthor{\bsnm{{Madsen}}, \binits{G.J.}},
\bauthor{\bsnm{{Chatterjee}}, \binits{S.}},
\bauthor{\bsnm{{Mao}}, \binits{S.A.}}:
\batitle{{The Vertical Structure of Warm Ionised Gas in the Milky Way}}.
\bjtitle{\pasa}
\bvolume{25},
\bfpage{184}--\blpage{200}
(\byear{2008})
DOI: \doiurl{10.1071/AS08004}
{arXiv:\href{https://arxiv.org/abs/0808.2550}{{0808.2550}}}
\end{barticle}
\endbibitem

%%% 17
\bibitem[\protect\citeauthoryear{Norman and Satoru}{1989}]{Norman1989}
\begin{barticle}
\bauthor{\bsnm{Norman}, \binits{C.A.}},
\bauthor{\bsnm{Satoru}, \binits{I.}}:
\batitle{{The Disk-Halo Interaction: Superbubbles and the Structure of the
  Interstellar Medium}}.
\bjtitle{\apj}
\bvolume{345},
\bfpage{372}
(\byear{1989})
DOI: \doiurl{10.1086/167912}
\end{barticle}
\endbibitem

%%% 18
\bibitem[\protect\citeauthoryear{Ponti et~al.}{2019}]{Ponti2019}
\begin{barticle}
\bauthor{\bsnm{Ponti}, \binits{G.}}, \betal:
\batitle{{An X-ray chimney extending hundreds of parsecs above and below the
  Galactic Centre}}.
\bjtitle{Nature}
\bvolume{567},
\bfpage{347}--\blpage{350}
(\byear{2019})
DOI: \doiurl{10.1038/s41586-019-1009-6}
{arXiv:\href{https://arxiv.org/abs/1904.05969}{{1904.05969}}}
\end{barticle}
\endbibitem

%%% 19
\bibitem[\protect\citeauthoryear{Aharonian et~al.}{2019}]{Aharonian2019}
\begin{barticle}
\bauthor{\bsnm{Aharonian}, \binits{F.}},
\bauthor{\bsnm{Yang}, \binits{R.}},
\bauthor{\bsnm{Wilhelmi}, \binits{E.}}:
\batitle{{Massive stars as major factories of Galactic cosmic rays}}.
\bjtitle{Nature Astronomy}
\bvolume{3},
\bfpage{561}--\blpage{567}
(\byear{2019})
DOI: \doiurl{10.1038/s41550-019-0724-0}
{arXiv:\href{https://arxiv.org/abs/1804.02331}{{1804.02331}}}
\end{barticle}
\endbibitem

%%% 20
\bibitem[\protect\citeauthoryear{Morlino et~al.}{2021}]{Morlino2021}
\begin{barticle}
\bauthor{\bsnm{Morlino}, \binits{G.}},
\bauthor{\bsnm{Blasi}, \binits{P.}},
\bauthor{\bsnm{Peretti}, \binits{E.}},
\bauthor{\bsnm{Cristofari}, \binits{P.}}:
\batitle{{Particle acceleration in winds of star clusters}}.
\bjtitle{\mnras}
\bvolume{504},
\bfpage{6096}--\blpage{6105}
(\byear{2021})
DOI: \doiurl{10.1093/mnras/stab690}
{arXiv:\href{https://arxiv.org/abs/2102.09217}{{2102.09217}}}
\end{barticle}
\endbibitem

%%% 21
\bibitem[\protect\citeauthoryear{Vieu and Reville}{2023}]{Vieu2023}
\begin{barticle}
\bauthor{\bsnm{Vieu}, \binits{T.}},
\bauthor{\bsnm{Reville}, \binits{B.}}:
\batitle{{Massive star cluster origin for the galactic cosmic ray population at
  very-high energies}}.
\bjtitle{\mnras}
\bvolume{519},
\bfpage{136}--\blpage{147}
(\byear{2023})
DOI: \doiurl{10.1093/mnras/stac3469}
{arXiv:\href{https://arxiv.org/abs/2211.11625}{{2211.11625}}}
\end{barticle}
\endbibitem

%%% 22
\bibitem[\protect\citeauthoryear{{Ackermann, M., \emph{et al.} (\fmlat
  Collaboration)}}{2011}]{FermiLAT_Cocoon_2011}
\begin{barticle}
\bauthor{\bsnm{{Ackermann, M., \emph{et al.} (\fmlat Collaboration)}}}:
\batitle{{A Cocoon of Freshly Accelerated Cosmic Rays Detected by Fermi in the
  Cygnus Superbubble}}.
\bjtitle{Science}
\bvolume{334},
\bfpage{1103}--\blpage{1107}
(\byear{2011})
DOI: \doiurl{10.1126/science.1210311}
\end{barticle}
\endbibitem

%%% 23
\bibitem[\protect\citeauthoryear{Yang and Aharonian}{2017}]{Yang2017}
\begin{barticle}
\bauthor{\bsnm{Yang}, \binits{R.-Z.}},
\bauthor{\bsnm{Aharonian}, \binits{F.}}:
\batitle{{Diffuse $\gamma$-ray emission near the young massive cluster NGC
  3603}}.
\bjtitle{\aap}
\bvolume{600},
\bfpage{107}
(\byear{2017})
DOI: \doiurl{10.1051/0004-6361/201630213}
{arXiv:\href{https://arxiv.org/abs/1612.02250}{{1612.02250}}}
\end{barticle}
\endbibitem

%%% 24
\bibitem[\protect\citeauthoryear{Yang and Wang}{2020}]{Yang2020}
\begin{barticle}
\bauthor{\bsnm{Yang}, \binits{R.-Z.}},
\bauthor{\bsnm{Wang}, \binits{Y.}}:
\batitle{{The diffuse gamma-ray emission toward the Galactic mini starburst
  W43}}.
\bjtitle{\aap}
\bvolume{640},
\bfpage{60}
(\byear{2020})
DOI: \doiurl{10.1051/0004-6361/202037518}
{arXiv:\href{https://arxiv.org/abs/2007.15295}{{2007.15295}}}
\end{barticle}
\endbibitem

%%% 25
\bibitem[\protect\citeauthoryear{Sun et~al.}{2020}]{Sun2020}
\begin{barticle}
\bauthor{\bsnm{Sun}, \binits{X.-N.}},
\bauthor{\bsnm{Yang}, \binits{R.-Z.}},
\bauthor{\bsnm{Wang}, \binits{X.-Y.}}:
\batitle{{Diffuse $\gamma$-ray emission from the vicinity of young massive star
  cluster RSGC 1}}.
\bjtitle{\mnras}
\bvolume{494},
\bfpage{3405}--\blpage{3412}
(\byear{2020})
DOI: \doiurl{10.1093/mnras/staa947}
{arXiv:\href{https://arxiv.org/abs/2006.02052}{{2006.02052}}}
\end{barticle}
\endbibitem

%%% 26
\bibitem[\protect\citeauthoryear{Liu et~al.}{2022}]{Liu2022}
\begin{barticle}
\bauthor{\bsnm{Liu}, \binits{B.}},
\bauthor{\bsnm{Yang}, \binits{R.-Z.}},
\bauthor{\bsnm{Chen}, \binits{Z.}}:
\batitle{{Gamma-ray observation towards the young massive star cluster NGC~6618
  in the M17 region}}.
\bjtitle{\mnras}
\bvolume{513},
\bfpage{4747}--\blpage{4753}
(\byear{2022})
DOI: \doiurl{10.1093/mnras/stac1252}
{arXiv:\href{https://arxiv.org/abs/2205.06430}{{2205.06430}}}
\end{barticle}
\endbibitem

%%% 27
\bibitem[\protect\citeauthoryear{{Peron} et~al.}{2024}]{Peron2024}
\begin{barticle}
\bauthor{\bsnm{{Peron}}, \binits{G.}},
\bauthor{\bsnm{{Casanova}}, \binits{S.}},
\bauthor{\bsnm{{Gabici}}, \binits{S.}},
\bauthor{\bsnm{{Baghmanyan}}, \binits{V.}},
\bauthor{\bsnm{{Aharonian}}, \binits{F.}}:
\batitle{{The contribution of winds from star clusters to the Galactic
  cosmic-ray population}}.
\bjtitle{Nature Astronomy}
\bvolume{8},
\bfpage{530}--\blpage{537}
(\byear{2024})
DOI: \doiurl{10.1038/s41550-023-02168-6}
{arXiv:\href{https://arxiv.org/abs/2407.07509}{{2407.07509}}}
\end{barticle}
\endbibitem

%%% 28
\bibitem[\protect\citeauthoryear{{Abramowski, A., \emph{et al.} (\hess
  Collaboration)}}{2011}]{HESS_Wd2_2011}
\begin{barticle}
\bauthor{\bsnm{{Abramowski, A., \emph{et al.} (\hess Collaboration)}}}:
\batitle{{Revisiting the Westerlund 2 field with the HESS telescope array}}.
\bjtitle{\aap}
\bvolume{525},
\bfpage{46}
(\byear{2011})
DOI: \doiurl{10.1051/0004-6361/201015290}
{arXiv:\href{https://arxiv.org/abs/1009.3012}{{1009.3012}}}
\end{barticle}
\endbibitem

%%% 29
\bibitem[\protect\citeauthoryear{{Abeysekara, A. U., \emph{et al.} (HAWC
  Collaboration)}}{2021}]{HAWC_Cygnus_2021}
\begin{barticle}
\bauthor{\bsnm{{Abeysekara, A. U., \emph{et al.} (HAWC Collaboration)}}}:
\batitle{{HAWC observations of the acceleration of very-high-energy cosmic rays
  in the Cygnus Cocoon}}.
\bjtitle{Nature Astronomy}
\bvolume{5},
\bfpage{465}--\blpage{471}
(\byear{2021})
DOI: \doiurl{10.1038/s41550-021-01318-y}
{arXiv:\href{https://arxiv.org/abs/2103.06820}{{2103.06820}}}
\end{barticle}
\endbibitem

%%% 30
\bibitem[\protect\citeauthoryear{{Aharonian, F., \emph{et al.} (\hess
  Collaboration)}}{2022}]{HESS_Wd1_2022}
\begin{barticle}
\bauthor{\bsnm{{Aharonian, F., \emph{et al.} (\hess Collaboration)}}}:
\batitle{{A deep spectromorphological study of the $\gamma$-ray emission
  surrounding the young massive stellar cluster Westerlund 1}}.
\bjtitle{\aap}
\bvolume{666},
\bfpage{124}
(\byear{2022})
DOI: \doiurl{10.1051/0004-6361/202244323}
{arXiv:\href{https://arxiv.org/abs/2207.10921}{{2207.10921}}}
\end{barticle}
\endbibitem

%%% 31
\bibitem[\protect\citeauthoryear{{Aharonian, F., \emph{et al.} (HESS
  Collaboration)}}{2024}]{HESS_LMC_2024}
\begin{barticle}
\bauthor{\bsnm{{Aharonian, F., \emph{et al.} (HESS Collaboration)}}}:
\batitle{{Very-high-energy $\gamma$-Ray Emission from Young Massive Star
  Clusters in the Large Magellanic Cloud}}.
\bjtitle{\apjl}
\bvolume{970},
\bfpage{21}
(\byear{2024})
DOI: \doiurl{10.3847/2041-8213/ad5e67}
{arXiv:\href{https://arxiv.org/abs/2407.16219}{{2407.16219}}}
\end{barticle}
\endbibitem

%%% 32
\bibitem[\protect\citeauthoryear{{Cao, Z., \emph{et al.} (LHAASO
  Collaboration)}}{2024}]{LHAASO_Cygnus_2024}
\begin{barticle}
\bauthor{\bsnm{{Cao, Z., \emph{et al.} (LHAASO Collaboration)}}}:
\batitle{{An ultrahigh-energy $\gamma$-ray bubble powered by a super
  PeVatron}}.
\bjtitle{Science Bulletin}
\bvolume{69},
\bfpage{449}--\blpage{457}
(\byear{2024})
DOI: \doiurl{10.1016/j.scib.2023.12.040}
{arXiv:\href{https://arxiv.org/abs/2310.10100}{{2310.10100}}}
\end{barticle}
\endbibitem

%%% 33
\bibitem[\protect\citeauthoryear{{Cao, Z., \emph{et al.} (LHAASO
  Collaboration)}}{2025}]{LHAASO_W43_2025}
\begin{barticle}
\bauthor{\bsnm{{Cao, Z., \emph{et al.} (LHAASO Collaboration)}}}:
\batitle{{Observation of the $\gamma$-ray emission from W43 with LHAASO}}.
\bjtitle{Science China Physics, Mechanics \& Astronomy}
\bvolume{68},
\bfpage{279502}
(\byear{2025})
DOI: \doiurl{10.1007/s11433-024-2477-9}
{arXiv:\href{https://arxiv.org/abs/2408.09905}{{2408.09905}}}
\end{barticle}
\endbibitem

%%% 34
\bibitem[\protect\citeauthoryear{Navarete et~al.}{2022}]{Navarete2022}
\begin{barticle}
\bauthor{\bsnm{Navarete}, \binits{F.}},
\bauthor{\bsnm{Damineli}, \binits{A.}},
\bauthor{\bsnm{Ramirez}, \binits{A.E.}},
\bauthor{\bsnm{Rocha}, \binits{D.}},
\bauthor{\bsnm{Almeida}, \binits{L.}}:
\batitle{{Distance and age of the massive stellar cluster Westerlund~1. I.
  Parallax method using Gaia-EDR3}}.
\bjtitle{\mnras}
\bvolume{516},
\bfpage{1289}--\blpage{1301}
(\byear{2022})
DOI: \doiurl{10.1093/mnras/stac2374}
{arXiv:\href{https://arxiv.org/abs/2204.09414}{{2204.09414}}}
\end{barticle}
\endbibitem

%%% 35
\bibitem[\protect\citeauthoryear{Negueruela et~al.}{2022}]{Negueruela2022}
\begin{barticle}
\bauthor{\bsnm{Negueruela}, \binits{I.}}, \betal:
\batitle{{Westerlund 1 under the light of Gaia EDR3: Distance, isolation,
  extent, and a hidden population}}.
\bjtitle{\aap}
\bvolume{664},
\bfpage{146}
(\byear{2022})
DOI: \doiurl{10.1051/0004-6361/202142985}
{arXiv:\href{https://arxiv.org/abs/2204.00422}{{2204.00422}}}
\end{barticle}
\endbibitem

%%% 36
\bibitem[\protect\citeauthoryear{Clark et~al.}{2005}]{Clark2005}
\begin{barticle}
\bauthor{\bsnm{Clark}, \binits{J.S.}},
\bauthor{\bsnm{Negueruela}, \binits{I.}},
\bauthor{\bsnm{Crowther}, \binits{P.A.}},
\bauthor{\bsnm{Goodwin}, \binits{S.P.}}:
\batitle{{On the massive stellar population of the super star cluster
  Westerlund 1}}.
\bjtitle{\aap}
\bvolume{434},
\bfpage{949}--\blpage{969}
(\byear{2005})
DOI: \doiurl{10.1051/0004-6361:20042413}
{arXiv:\href{https://arxiv.org/abs/astro-ph/0504342}{{astro-ph/0504342}}}
\end{barticle}
\endbibitem

%%% 37
\bibitem[\protect\citeauthoryear{Crowther et~al.}{2006}]{Crowther2006}
\begin{barticle}
\bauthor{\bsnm{Crowther}, \binits{P.A.}},
\bauthor{\bsnm{Hadfield}, \binits{L.J.}},
\bauthor{\bsnm{Clark}, \binits{J.S.}},
\bauthor{\bsnm{Negueruela}, \binits{I.}},
\bauthor{\bsnm{Vacca}, \binits{W.D.}}:
\batitle{{A census of the Wolf–Rayet content in Westerlund 1 from
  near-infrared imaging and spectroscopy}}.
\bjtitle{\mnras}
\bvolume{372},
\bfpage{1407}--\blpage{1424}
(\byear{2006})
DOI: \doiurl{10.1111/j.1365-2966.2006.10952.x}
{arXiv:\href{https://arxiv.org/abs/astro-ph/0608356}{{astro-ph/0608356}}}
\end{barticle}
\endbibitem

%%% 38
\bibitem[\protect\citeauthoryear{Brandner et~al.}{2008}]{Brandner2008}
\begin{barticle}
\bauthor{\bsnm{Brandner}, \binits{W.}}, \betal:
\batitle{{Intermediate to low-mass stellar content of Westerlund 1}}.
\bjtitle{\aap}
\bvolume{478},
\bfpage{137}--\blpage{149}
(\byear{2008})
DOI: \doiurl{10.1051/0004-6361:20077579}
{arXiv:\href{https://arxiv.org/abs/0711.1624}{{0711.1624}}}
\end{barticle}
\endbibitem

%%% 39
\bibitem[\protect\citeauthoryear{Beasor et~al.}{2021}]{Beasor2021}
\begin{barticle}
\bauthor{\bsnm{Beasor}, \binits{E.R.}},
\bauthor{\bsnm{Davies}, \binits{B.}},
\bauthor{\bsnm{Smith}, \binits{N.}},
\bauthor{\bsnm{Gehrz}, \binits{R.D.}},
\bauthor{\bsnm{Figer}, \binits{D.F.}}:
\batitle{{The Age of Westerland 1 Revisited}}.
\bjtitle{\apj}
\bvolume{912},
\bfpage{16}
(\byear{2021})
DOI: \doiurl{10.3847/1538-4357/abec44}
{arXiv:\href{https://arxiv.org/abs/2103.02609}{{2103.02609}}}
\end{barticle}
\endbibitem

%%% 40
\bibitem[\protect\citeauthoryear{Muno et~al.}{2006}]{Muno2006}
\begin{barticle}
\bauthor{\bsnm{Muno}, \binits{M.P.}}, \betal:
\batitle{{Diffuse, Nonthermal X-Ray Emission from the Galactic Star Cluster
  Westerlund 1}}.
\bjtitle{\apj}
\bvolume{650},
\bfpage{203}
(\byear{2006})
DOI: \doiurl{10.1086/507175}
{arXiv:\href{https://arxiv.org/abs/astro-ph/0606492}{{astro-ph/0606492}}}
\end{barticle}
\endbibitem

%%% 41
\bibitem[\protect\citeauthoryear{H\"arer et~al.}{2023}]{Haerer2023}
\begin{barticle}
\bauthor{\bsnm{H\"arer}, \binits{L.K.}},
\bauthor{\bsnm{Reville}, \binits{B.}},
\bauthor{\bsnm{Hinton}, \binits{J.}},
\bauthor{\bsnm{Mohrmann}, \binits{L.}},
\bauthor{\bsnm{Vieu}, \binits{T.}}:
\batitle{{Understanding the TeV $\gamma$-ray emission surrounding the young
  massive star cluster Westerlund 1}}.
\bjtitle{\aap}
\bvolume{671},
\bfpage{4}
(\byear{2023})
DOI: \doiurl{10.1051/0004-6361/202245444}
{arXiv:\href{https://arxiv.org/abs/2301.10496}{{2301.10496}}}
\end{barticle}
\endbibitem

%%% 42
\bibitem[\protect\citeauthoryear{{Atwood, W. B., \emph{et al.} (\fmlat
  Collaboration)}}{2009}]{FermiLAT2009}
\begin{barticle}
\bauthor{\bsnm{{Atwood, W. B., \emph{et al.} (\fmlat Collaboration)}}}:
\batitle{{The Large Area Telescope on the Fermi Gamma-ray Space Telescope
  Mission}}.
\bjtitle{\apj}
\bvolume{697},
\bfpage{1071}
(\byear{2009})
DOI: \doiurl{10.1088/0004-637X/697/2/1071}
{arXiv:\href{https://arxiv.org/abs/0902.1089}{{0902.1089}}}
\end{barticle}
\endbibitem

%%% 43
\bibitem[\protect\citeauthoryear{{Baumgartner} and
  {Breitschwerdt}}{2013}]{Baumgartner}
\begin{barticle}
\bauthor{\bsnm{{Baumgartner}}, \binits{V.}},
\bauthor{\bsnm{{Breitschwerdt}}, \binits{D.}}:
\batitle{{Superbubble evolution in disk galaxies. I. Study of blow-out by
  analytical models}}.
\bjtitle{\aap}
\bvolume{557},
\bfpage{140}
(\byear{2013})
DOI: \doiurl{10.1051/0004-6361/201321261}
{arXiv:\href{https://arxiv.org/abs/1402.0194}{{1402.0194}}}
\end{barticle}
\endbibitem

%%% 44
\bibitem[\protect\citeauthoryear{{Kalberla} and {Haud}}{2015}]{Kalberla2015}
\begin{barticle}
\bauthor{\bsnm{{Kalberla}}, \binits{P.M.W.}},
\bauthor{\bsnm{{Haud}}, \binits{U.}}:
\batitle{{GASS: The Parkes Galactic All-Sky Survey. Update: improved correction
  for instrumental effects and new data release}}.
\bjtitle{\aap}
\bvolume{578},
\bfpage{78}
(\byear{2015})
DOI: \doiurl{10.1051/0004-6361/201525859}
{arXiv:\href{https://arxiv.org/abs/1505.01011}{{1505.01011}}}
\end{barticle}
\endbibitem

%%% 45
\bibitem[\protect\citeauthoryear{Hahn}{2015}]{Hahn2015}
\begin{bchapter}
\bauthor{\bsnm{Hahn}, \binits{J.}}:
\bctitle{{GAMERA - A Modular Framework For Spectral Modeling In VHE
  Astronomy}}.
In: \bbtitle{Proc. 34th Int. Cosmic Ray Conf. (ICRC2015)},
p. \bfpage{917}
(\byear{2015}).
DOI: \doiurl{10.22323/1.236.0917}
\end{bchapter}
\endbibitem

%%% 46
\bibitem[\protect\citeauthoryear{{Hahn} et~al.}{2022}]{Hahn2020}
\begin{botherref}
\oauthor{\bsnm{{Hahn}}, \binits{J.}},
\oauthor{\bsnm{{Romoli}}, \binits{C.}},
\oauthor{\bsnm{{Breuhaus}}, \binits{M.}}:
{GAMERA: Source modeling in gamma astronomy}.
Astrophysics Source Code Library
\textbf{ascl:2203.007}
(2022).
\url{https://ascl.net/2203.007}
\end{botherref}
\endbibitem

%%% 47
\bibitem[\protect\citeauthoryear{{Abdollahi, S., \emph{et al.} (\fmlat
  Collaboration)}}{2020}]{FermiLAT_4FGL_2020}
\begin{barticle}
\bauthor{\bsnm{{Abdollahi, S., \emph{et al.} (\fmlat Collaboration)}}}:
\batitle{{Fermi Large Area Telescope Fourth Source Catalog}}.
\bjtitle{\apjs}
\bvolume{247},
\bfpage{33}
(\byear{2020})
DOI: \doiurl{10.3847/1538-4365/ab6bcb}
{arXiv:\href{https://arxiv.org/abs/1902.10045}{{1902.10045}}}
\end{barticle}
\endbibitem

%%% 48
\bibitem[\protect\citeauthoryear{{Abdollahi, S., \emph{et al.} (\fmlat
  Collaboration)}}{2022}]{FermiLAT_4FGLDR3_2022}
\begin{barticle}
\bauthor{\bsnm{{Abdollahi, S., \emph{et al.} (\fmlat Collaboration)}}}:
\batitle{{Incremental Fermi Large Area Telescope Fourth Source Catalog}}.
\bjtitle{\apjs}
\bvolume{260},
\bfpage{53}
(\byear{2022})
DOI: \doiurl{10.3847/1538-4365/ac6751}
{arXiv:\href{https://arxiv.org/abs/2201.11184}{{2201.11184}}}
\end{barticle}
\endbibitem

%%% 49
\bibitem[\protect\citeauthoryear{{Ballet} et~al.}{2023}]{FermiLAT_4FGLDR4_2023}
\begin{botherref}
\oauthor{\bsnm{{Ballet}}, \binits{J.}},
\oauthor{\bsnm{{Bruel}}, \binits{P.}},
\oauthor{\bsnm{{Burnett}}, \binits{T.H.}},
\oauthor{\bsnm{{Lott}}, \binits{B.}},
\oauthor{\bsnm{{The Fermi-LAT collaboration}}}:
{Fermi Large Area Telescope Fourth Source Catalog Data Release 4 (4FGL-DR4)}.
arXiv e-prints
(2023)
{arXiv:\href{https://arxiv.org/abs/2307.12546}{{2307.12546}}}
\end{botherref}
\endbibitem

%%% 50
\bibitem[\protect\citeauthoryear{Akaike}{1974}]{Akaike1974}
\begin{barticle}
\bauthor{\bsnm{Akaike}, \binits{H.}}:
\batitle{{A new look at the statistical model identification}}.
\bjtitle{IEEE Transactions on Automatic Control}
\bvolume{19},
\bfpage{716}--\blpage{723}
(\byear{1974})
DOI: \doiurl{10.1109/TAC.1974.1100705}
\end{barticle}
\endbibitem

%%% 51
\bibitem[\protect\citeauthoryear{{Manchester} et~al.}{2005}]{Manchester2005}
\begin{barticle}
\bauthor{\bsnm{{Manchester}}, \binits{R.N.}},
\bauthor{\bsnm{{Hobbs}}, \binits{G.B.}},
\bauthor{\bsnm{{Teoh}}, \binits{A.}},
\bauthor{\bsnm{{Hobbs}}, \binits{M.}}:
\batitle{{The Australia Telescope National Facility Pulsar Catalogue}}.
\bjtitle{\aj}
\bvolume{129},
\bfpage{1993}--\blpage{2006}
(\byear{2005})
DOI: \doiurl{10.1086/428488}
{arXiv:\href{https://arxiv.org/abs/astro-ph/0412641}{{astro-ph/0412641}}}
\end{barticle}
\endbibitem

%%% 52
\bibitem[\protect\citeauthoryear{{Ackermann, M., \emph{et al.} (\fmlat
  Collaboration)}}{2017}]{FermiLAT_FGES_2017}
\begin{barticle}
\bauthor{\bsnm{{Ackermann, M., \emph{et al.} (\fmlat Collaboration)}}}:
\batitle{{Search for Extended Sources in the Galactic Plane Using Six Years of
  Fermi-Large Area Telescope Pass 8 Data above 10 GeV}}.
\bjtitle{\apj}
\bvolume{843},
\bfpage{139}
(\byear{2017})
DOI: \doiurl{10.3847/1538-4357/aa775a}
{arXiv:\href{https://arxiv.org/abs/1702.00476}{{1702.00476}}}
\end{barticle}
\endbibitem

%%% 53
\bibitem[\protect\citeauthoryear{Ohm et~al.}{2013}]{Ohm2013}
\begin{barticle}
\bauthor{\bsnm{Ohm}, \binits{S.}},
\bauthor{\bsnm{Hinton}, \binits{J.A.}},
\bauthor{\bsnm{White}, \binits{R.}}:
\batitle{{$\gamma$-ray emission from the Westerlund 1 region}}.
\bjtitle{\mnras}
\bvolume{434},
\bfpage{2289}--\blpage{2294}
(\byear{2013})
DOI: \doiurl{10.1093/mnras/stt1170}
{arXiv:\href{https://arxiv.org/abs/1306.5642}{{1306.5642}}}
\end{barticle}
\endbibitem

%%% 54
\bibitem[\protect\citeauthoryear{Bruel}{2021}]{Bruel2021}
\begin{barticle}
\bauthor{\bsnm{Bruel}, \binits{P.}}:
\batitle{{A new method to perform data-model comparison in \emph{Fermi}-LAT
  analysis}}.
\bjtitle{\aap}
\bvolume{656},
\bfpage{81}
(\byear{2021})
DOI: \doiurl{10.1051/0004-6361/202141553}
{arXiv:\href{https://arxiv.org/abs/2109.07443}{{2109.07443}}}
\end{barticle}
\endbibitem

%%% 55
\bibitem[\protect\citeauthoryear{Haubner et~al.}{2025}]{Haubner2025}
\begin{barticle}
\bauthor{\bsnm{Haubner}, \binits{K.}}, \betal:
\batitle{{eROSITA X-ray analysis of the PeVatron candidate Westerlund 1}}.
\bjtitle{\aap}
\bvolume{695},
\bfpage{3}
(\byear{2025})
DOI: \doiurl{10.1051/0004-6361/202451964}
{arXiv:\href{https://arxiv.org/abs/2501.12990}{{2501.12990}}}
\end{barticle}
\endbibitem

%%% 56
\bibitem[\protect\citeauthoryear{{Reid} et~al.}{2019}]{Reid2019}
\begin{barticle}
\bauthor{\bsnm{{Reid}}, \binits{M.J.}}, \betal:
\batitle{{Trigonometric Parallaxes of High-mass Star-forming Regions: Our View
  of the Milky Way}}.
\bjtitle{\apj}
\bvolume{885},
\bfpage{131}
(\byear{2019})
DOI: \doiurl{10.3847/1538-4357/ab4a11}
{arXiv:\href{https://arxiv.org/abs/1910.03357}{{1910.03357}}}
\end{barticle}
\endbibitem

%%% 57
\bibitem[\protect\citeauthoryear{Roman-Duval et~al.}{2009}]{RomanDuval2009}
\begin{barticle}
\bauthor{\bsnm{Roman-Duval}, \binits{J.}}, \betal:
\batitle{{Kinematic Distances to Molecular Clouds identified in the Galactic
  Ring Survey}}.
\bjtitle{\apj}
\bvolume{699},
\bfpage{1153}
(\byear{2009})
DOI: \doiurl{10.1088/0004-637X/699/2/1153}
{arXiv:\href{https://arxiv.org/abs/0905.0723}{{0905.0723}}}
\end{barticle}
\endbibitem

%%% 58
\bibitem[\protect\citeauthoryear{{Abuter, R., \emph{et al.} (GRAVITY
  Collaboration)}}{2019}]{GRAVITY2019}
\begin{barticle}
\bauthor{\bsnm{{Abuter, R., \emph{et al.} (GRAVITY Collaboration)}}}:
\batitle{{A geometric distance measurement to the Galactic center black hole
  with 0.3\% uncertainty}}.
\bjtitle{\aap}
\bvolume{625},
\bfpage{10}
(\byear{2019})
DOI: \doiurl{10.1051/0004-6361/201935656}
{arXiv:\href{https://arxiv.org/abs/1904.05721}{{1904.05721}}}
\end{barticle}
\endbibitem

%%% 59
\bibitem[\protect\citeauthoryear{{Russeil}}{2003}]{Russeil2003}
\begin{barticle}
\bauthor{\bsnm{{Russeil}}, \binits{D.}}:
\batitle{{Star-forming complexes and the spiral structure of our Galaxy}}.
\bjtitle{\aap}
\bvolume{397},
\bfpage{133}--\blpage{146}
(\byear{2003})
DOI: \doiurl{10.1051/0004-6361:20021504}
\end{barticle}
\endbibitem

%%% 60
\bibitem[\protect\citeauthoryear{Fok et~al.}{2012}]{Fok2012}
\begin{barticle}
\bauthor{\bsnm{Fok}, \binits{T.K.T.}},
\bauthor{\bsnm{Nakashima}, \binits{J.}},
\bauthor{\bsnm{Yung}, \binits{B.H.K.}},
\bauthor{\bsnm{Hsia}, \binits{C.}},
\bauthor{\bsnm{Deguchi}, \binits{S.}}:
\batitle{{Maser Observations of Westerlund 1 and Comprehensive Considerations
  on Maser Properties of Red Supergiants Associated with Massive Clusters}}.
\bjtitle{\apj}
\bvolume{760},
\bfpage{65}
(\byear{2012})
DOI: \doiurl{10.1088/0004-637X/760/1/65}
{arXiv:\href{https://arxiv.org/abs/1209.6427}{{1209.6427}}}
\end{barticle}
\endbibitem

%%% 61
\bibitem[\protect\citeauthoryear{Kothes and Dougherty}{2007}]{Kothes2007}
\begin{barticle}
\bauthor{\bsnm{Kothes}, \binits{R.}},
\bauthor{\bsnm{Dougherty}, \binits{S.M.}}:
\batitle{{The distance and neutral environment of the massive stellar cluster
  Westerlund 1}}.
\bjtitle{\aap}
\bvolume{468},
\bfpage{993}--\blpage{1000}
(\byear{2007})
DOI: \doiurl{10.1051/0004-6361:20077309}
{arXiv:\href{https://arxiv.org/abs/0704.3073}{{0704.3073}}}
\end{barticle}
\endbibitem

%%% 62
\bibitem[\protect\citeauthoryear{{Hou}}{2021}]{Hou2021}
\begin{barticle}
\bauthor{\bsnm{{Hou}}, \binits{L.G.}}:
\batitle{{The spiral structure in the Solar neighbourhood}}.
\bjtitle{Frontiers in Astronomy and Space Sciences}
\bvolume{8},
\bfpage{103}
(\byear{2021})
DOI: \doiurl{10.3389/fspas.2021.671670}
{arXiv:\href{https://arxiv.org/abs/2110.04446}{{2110.04446}}}
\end{barticle}
\endbibitem

%%% 63
\bibitem[\protect\citeauthoryear{{Popescu} et~al.}{2017}]{Popescu2017}
\begin{barticle}
\bauthor{\bsnm{{Popescu}}, \binits{C.C.}}, \betal:
\batitle{{A radiation transfer model for the Milky Way: I. Radiation fields and
  application to high-energy astrophysics}}.
\bjtitle{\mnras}
\bvolume{470},
\bfpage{2539}--\blpage{2558}
(\byear{2017})
DOI: \doiurl{10.1093/mnras/stx1282}
{arXiv:\href{https://arxiv.org/abs/1705.06652}{{1705.06652}}}
\end{barticle}
\endbibitem

\end{thebibliography}
\end{document}